\documentclass[usenatbib]{mn2e}

\usepackage{graphics}
\usepackage{epsfig}
\usepackage{natbib}

\newcommand{\kms}{km~s\ensuremath{^{-1}}}
\newcommand{\msun}{$M_{\odot}$}
\newcommand{\lsun}{$L_{\odot}$}
\newcommand{\nuvr}{NUV$-r$}
\newcommand{\mh}{$M_{\rm H_2}$}
\newcommand{\mhi}{$M_{\rm HI}$}
\newcommand{\hi}{{H{\sc I}}}
\newcommand{\mstar}{$M_{\ast}$}
\newcommand{\must}{$\mu_{\ast}$}
\newcommand{\hmol}{H$_2$}
\newcommand{\sfrsed}{SFR$_{SED}$}
\newcommand{\sfrtot}{SFR$_{tot}$}
\newcommand{\tdep}{$t_{dep}(\rm H_2)$}
\newcommand{\tdepHI}{$t_{dep}({\rm \hi})$}
\newcommand{\paperI}{Paper I}

\newcommand{\ntot}{222}

\newcommand{\nspec}{113}

\newcommand{\nodata}{$...$}

\newcommand{\apj}{ApJ}
\newcommand{\apjs}{ApJS}
\newcommand{\apjl}{ApJ}
\newcommand{\aj}{AJ}
\newcommand{\mnras}{MNRAS}
\newcommand{\aap}{A\&A}

\newcommand{\araa}{ARA\&A}

\newcommand{\nat}{Nature}

\begin{document}

\title[COLD GASS: Variable Molecular Gas Depletion Times]{COLD GASS, an IRAM Legacy Survey of Molecular Gas in Massive Galaxies:  II.  The non-universality of the Molecular Gas Depletion Timescale}

\author[A. Saintonge et al.]{Am\'{e}lie Saintonge$^{1,2}$\thanks{E-mail: amelie@mpe.mpg.de}, Guinevere Kauffmann$^{1}$, Jing Wang$^{1,3}$, Carsten Kramer$^{4}$,   
\newauthor
Linda J. Tacconi$^{2}$, Christof Buchbender$^{4}$, Barbara Catinella$^{1}$,  Javier Graci\'{a}-Carpio$^{2}$, 
\newauthor
Luca Cortese$^{5}$, Silvia Fabello$^{1}$, Jian Fu$^{6,1}$, Reinhard Genzel$^{2}$, Riccardo Giovanelli$^{7}$, 
\newauthor
Qi Guo$^{8,9}$, Martha P. Haynes$^{7}$, Timothy M. Heckman$^{10}$, Mark R. Krumholz$^{11}$, 
\newauthor
Jenna Lemonias$^{12}$, Cheng Li$^{6,13}$,  Sean Moran$^{10}$, Nemesio Rodriguez-Fernandez$^{14}$, 
\newauthor
David Schiminovich$^{12}$, Karl Schuster$^{14}$ and Albrecht Sievers$^{4}$ \\
$^{1}$Max-Planck Institut f\"{u}r Astrophysik, 85741 Garching, Germany\\
$^{2}$Max-Planck Institut f\"{u}r extraterrestrische Physik, 85741 Garching, Germany\\
$^{3}$Center for Astrophysics, University of Science and Technology of China, 230026 Hefei, China\\
$^{4}$Instituto Radioastronom\'{i}a Milim\'{e}trica, Av. Divina Pastora 7, Nucleo Central, 18012 Granada, Spain\\
$^{5}$European Southern Observatory, Karl-Schwarzschild-Str. 2, 85748 Garching, Germany\\
$^{6}$Key Laboratory for Research in Galaxies and Cosmology, Shanghai Astronomical Observatory, Chinese Academy of Sciences, \\ 
Nandan Road 80, Shanghai 200030, China\\
$^{7}$Center for Radiophysics and Space Research, Cornell University, Ithaca, NY 14853, USA\\
$^{8}$National Astronomical Observatories, Chinese Academy of Sciences, Beijing 100012, China\\
$^{9}$Institute for Computational Cosmology, Department of Physics, Durham University, South Road, Durham DH1 3LE, UK\\
$^{10}$Johns Hopkins University, Baltimore, Maryland 21218, USA\\
$^{11}$Department of Astronomy and Astrophysics, University of California, Santa Cruz, CA 95064, USA\\
$^{12}$Department of Astronomy, Columbia University, New York, NY 10027, USA\\
$^{13}$Max-Planck-Institut Partner Group, Shanghai Astronomical Observatory\\
$^{14}$Institut de Radioastronomie Millim\'{e}trique, 300 Rue de la piscine, 38406 St Martin d'H\`{e}res, France}

\maketitle

\begin{abstract} 
We study the relation between molecular gas and star formation in a
volume-limited sample of \ntot\ galaxies from the COLD GASS survey,  
with measurements of the CO(1-0) line from the IRAM 30m telescope. The galaxies are
at redshifts $0.025<z<0.05$ and have stellar masses in the range
$10.0<\log M_{\ast}/M_{\odot}<11.5$.  The IRAM measurements are
complemented by deep Arecibo HI observations and homogeneous SDSS and
GALEX photometry. A reference sample that includes both UV and far-IR
data is used to calibrate our estimates of star formation rates
from the seven optical/UV bands.  The mean molecular gas depletion timescale (\tdep)
for all the galaxies in our sample is 1 Gyr, however  \tdep\ increases
by a factor of 6 from a value of $\sim 0.5$ Gyr for galaxies with stellar masses
of $\sim 10^{10} M_{\odot}$  to $\sim 3$ Gyr for galaxies with masses of a
few $\times 10^{11} M_{\odot}$.  In contrast, the atomic gas depletion
timescale remains contant at a value of around 3 Gyr.  This implies that
in high mass galaxies,  molecular and atomic gas depletion timescales are
comparable, but in low mass galaxies, molecular gas is being consumed much
more quickly than atomic gas.  The strongest dependences of \tdep\ are on
the stellar mass of the galaxy (parameterized
as $\log$ \tdep$= (0.36\pm0.07)(\log M_{\ast} - 10.70)+(9.03\pm0.99)$), and on the 
specific star formation rate. A single  \tdep\ versus sSFR relation
is able to fit both ``normal'' star-forming galaxies in our COLD GASS
sample, as well as   more extreme starburst galaxies (LIRGs and ULIRGs),
which have \tdep\ $< 10^8$ yr.  Normal galaxies at z=1-2 are displaced
with respect to the local galaxy population in the \tdep\ versus sSFR
plane and have molecular gas depletion times that are a factor of 3-5
times longer at a given value of sSFR due to their significantly larger gas fractions. 
\end{abstract}

\begin{keywords}
galaxies: fundamental parameters -- galaxies: evolution -- galaxies: ISM -- radio lines: galaxies -- surveys
\end{keywords}

\section{Introduction}

The physics governing the formation of stars is highly non-linear
and operates over a vast range in physical scale.  The size of the
self-gravitating cores of molecular clouds in which individual stars form
is only $\sim 0.1$ pc and the evolution of these cores is governed by
an array of highly complex physical processes: turbulence and  magnetic
fields in the interstellar medium may regulate the rate of collapse of
these structures, while winds, jet-like outflows, radiation pressure
and ionizing radiation from the most massive stars will act to
suppress the formation of young stars \citep[see][for a recent review]{mckee07}.

It is therefore not surprising that the rate at which gas is being
converted to stars in a single galaxy exhibits large scatter from one 
region to another when studied on small scales
\citep[e.g.][]{schruba10}. What is perhaps more remarkable
is the fact that once the smoothing scale exceeds a radius of $\sim 1$
kpc, well-defined relations between star formation and the {\em global}
gaseous properties of galaxies begin to appear. 

The first study of large-scale star formation in galaxies was by \citet{schmidt59},
 who studied the relative distribution of atomic gas and young
stars perpendicular to the galactic plane, and derived a power-law
relation between the rate of star formation and the volume density of
interstellar gas with exponent $n \sim 2$. Almost all studies since then
have compared the surface density of newly-formed 
stars with that of the total
(atomic+molecular) cold gas
\citep[e.g.][]{kennicutt89,kennicutt98a,kennicutt07,bigiel08}. These studies 
have confirmed the existence of a power-law relation of the form $\Sigma_{\rm
SFR}=a \Sigma^N_{\rm gas}$, with N generally in the range 1.3.-1.4.

It should be noted, however, that the exponent in the power law is
quite sensitive to the gaseous tracer that is used, as well as to the
type of galaxy under investigation and the size scale being considered. The steepest relations are generally
obtained for lower mass disk galaxies in which atomic hydrogen dominates
the total gas surface density \citep{kennicutt89,leroy08}. In the
inner disks of the more massive nearby spirals, molecular gas dominates. In
recent work that made use of high-resolution HI, CO, 24$\mu$m and UV maps
of nearby spiral galaxies, \citet{bigiel08} found that $\Sigma_{\rm
SFR}$ and $\Sigma_{H2}$ were related by a power law relation with $N=1$,
i.e. they found that $H_2$ appeared to form stars at a constant efficiency.
The range of $\Sigma_{\rm H2}$ over which this
relation was applicable was between 3 and 50 $M_{\odot}$ pc$^{-1}$,
i.e. significantly lower than gas densities 
in starburst galaxies. In samples of 
infrared-selected galaxies, including ultra-luminous infrared galaxies
with $L_{IR} > 10^{12} L_{\odot}$, there 
is an almost linear relation between star formation rate and the total amount of
very dense molecular gas traced by HCN rather than by CO \citep{gao04,gracia08}. 
Taken together, these results may imply
that the amount of dense gas available for star formation is higher
in more actively star-forming galaxies, but that SFR in the {\em very
densest phases} may be independent of global galaxy poperties \citep{mckee07}.

Most of these results are, however, based on the study of single or of 
relatively small samples of objects with similar  properties
(either nearby star-forming spiral galaxies or IR-selected
samples of starbursting systems).  Recently, \citet{GASS2} studied
the relation between atomic gas content and star formation
using an unbiased volume-limited sample of galaxies
with $M_* > 10^{10} M_{\odot}$ from
the {\it GALEX} Arecibo SDSS Survey (GASS).  They found an average HI
gas consumption timescale of $\sim3$ Gyr. Although the galaxy-to-galaxy
scatter around this average value was  large, they did not find any
dependence of the efficiency with which HI is converted into stars on
parameters such as stellar mass, stellar mass surface density, colour,
and concentration index.  This result is perhaps not surprising, since
the atomic gas is one step removed from the actual star formation process.

The relation between atomic gas, $H_2$ and star formation was also
extensively investigated by the THINGS/HERACLES team in a series
of papers \citep{THINGS,leroy08,bigiel08,leroy09}.
 In their sample, the HI-to-$H_2$ ratio varies with radius,
stellar surface density and pressure.  These authors suggested that
mechanisms occurring at the GMC scale were responsible for these
trends \citep{leroy08}.  Another possible scenario is that some fraction of
the atomic gas in galaxies should be regarded as 
 repository of baryonic material that
has been accreted from the surrounding intergalactic medium. Theory
predicts that this material 
resides predominantly in the outer regions of galaxies
\citep[e.g.][]{MMW98} and much of it therefore has
not yet reached high enough densities for self-shielding from the ambient
UV field to permit molecule formation to take place \citep[e.g.][]{fu10}.

Once molecular gas is formed, however, the picture painted by the
THINGS/HERACLES analyses of normal spirals is very simple:          
one would expect star formation to occur at a rate that
would deplete the $H_2$ in a universal timescale of $\sim2$ Gyr
\citep{leroy08,bigiel08}. Intriguingly, the same universal timescale also appears
to hold for the ``normal'' population of star-forming disk
galaxies at redshifts $1<z<2$ \citep{genzel10}.

We are conducting the COLD GASS survey, which will obtain
CO(1-0) line measurements and hence the molecular gas
content of a sample of $\sim 350$ massive
galaxies ($10.0<\log M_{\ast}/M_{\odot}<11.5$) using the IRAM 30m
telescope.  The details of the survey, data for the first half of the sample, and first science results were
presented in \citet{COLDGASS1} (hereafter, \paperI).  Since the COLD GASS
sample is selected only by stellar mass, it enables us to study the 
relation between molecular gas and star formation for a 
representative sample of galaxies, and to test
whether the universal depletion timescale that is observed to hold
when the data is smoothed on scales of a few hundred parsecs   
in small samples of nearby spirals, is truly applicable  
to the galaxy population at large.

In Section \ref{data} we describe the COLD GASS sample of local massive galaxies and present the
different data sets and measurements used in this paper.  In Section
\ref{results} we present our main result concerning the non-universality
of the molecular gas depletion timescale, which we summarize in Section \ref{summary} and discuss in
Section \ref{discussion}.  Finally,  in Section \ref{highz}
we compare our results for the large and unbiased
COLD GASS sample with results from the  literature on molecular
gas depletion timescales in normal and starburst galaxies, 
both at low and at high redshifts.
Throughout the paper, distance-dependent quantities are calculated for
a standard flat $\Lambda$CDM cosmology with $H_0=70$\kms\ Mpc$^{-1}$,
and we adopt a conversion factor from CO luminosity to $H_2$ mass of
$\alpha_{CO}=3.2$\msun\ (K \kms\ pc$^2$)$^{-1}$ (which does not account
for the presence of Helium), unless otherwise specified.

\section{Data}
\label{data}

The COLD GASS sample is stellar mass-selected ($M_{\ast}>10^{10}$\msun) 
and volume-limited ($0.025<z<0.050$).  The galaxies are selected from
the area of sky covered by the SDSS, the GALEX Medium Imaging 
Survey (MIS) and the ALFALFA HI survey.  This provides us with optical 
imaging and spectroscopy (over the central 3\arcsec) for all galaxies
in our sample, as well as HI fluxes for
the most gas-rich systems.  For the remaining galaxies, 
the HI data is obtained through pointed observations at the 
Arecibo observatory as part of the GASS survey \citep{GASS1}. 
The sample used here contains the \ntot\ galaxies presented in \paperI. 
The sample selection, observations and data products are fully described
in that paper, but we
present a short recapitulation here. 

\subsection{Optical and UV photometry}

Parameters such as redshifts, sizes, magnitudes, and Galactic extinction 
factors are retrieved from the  SDSS DR7 database \citep{DR7}.  
The UV data are taken from the {\it GALEX} All-sky and Medium Imaging 
surveys \citep[AIS and MIS, respectively, see][]{martin05}. 
The SDSS and {\it GALEX} images are, however, reprocessed following 
\citet{wang10}, in order to obtain accurate UV/optical matched-aperture 
photometry.  This process includes accurately registering the GALEX and SDSS images 
and smoothing them so that they have the same resolution.  
In practice, the SDSS $r-$band images are convolved to the resolution 
of the GALEX UV images before Sextractor is used to calculate magnitudes 
in consistent apertures,  therefore ensuring that measurements in different 
bands cover the same physical regions of the galaxies.  The derived 
\nuvr\ colours are corrected for Galactic extinction using the 
prescription of \citet{wyder07} \citep[see also][]{GASS1}. 
 Stellar masses are calculated from the SDSS photometry using the 
technique of \citet{salim07} assuming a Chabrier IMF.  The optical- and 
UV-derived quantities used throughout this study are published in
Table 1 of \paperI. 

In this study, we also consider morphological parameters derived 
for the GASS sample by \citet{GASS3}, following the technique 
of \citet{lotz04}. The asymmetry parameter (Asym) provides a measure 
of the difference between the SDSS $g-$band image of a galaxy 
and a copy of the same image, rotated by 180$^{\circ}$. 
The larger the value of Asym, the more asymmetric the galaxy.  
The other parameter we investigate in our analysis is 
smoothness (Smth), computed from the difference between 
the original $g-$band image and a  
copy of the same image smoothed to a scale of 20\% of the petrosian radius. 
Galaxies with clumpier morphologies therefore have larger values of Smth.  

\subsection{Optical spectroscopy}
\label{spectra}

For \nspec\ of the COLD GASS galaxies in the present sample, long-slit 
optical spectra were obtained with the MMT 6.5m and the Apache Point 
Observatory (APO) 3.5 m telescopes.  The spectra were taken along the 
major axis of the galaxies and cover the wavelength range 3900-7000 \AA. 
Data reduction includes standard procedures: biasing, flat-fielding, 
cosmic ray rejection, background estimation, and finally registration 
and co-addition of the individual frames.  Flux calibration is 
achieved through observations of spectrophotometric standard stars 
and by bootstraping to SDSS photometry measured through equivalent 
apertures.  These final calibrated spectra are then fitted with a family 
of templates derived from \citet{bc03} stellar population models 
following the approach of Tremonti et al (2004) and \citet{brinchmann04}. 
After subtraction of the stellar continuum, the strength of key 
emission features is measured by Gaussian fitting and the emission
line fluxes are corrected for dust extinction using the measured 
Balmer decrement. These procedures are described in detail in \citet{moran10}.

\subsection{Arecibo observations}

The HI observations are described in detail in \citet{GASS1}, so we 
only provide a brief overview here. The GASS survey builds upon existing 
HI databases: the Cornell digital HI archive \citep{springob05} and 
the ALFALFA survey \citep{ALFALFA1}.  HI data for about 20\% of the 
GASS sample (the most gas-rich objects), can be found in either of these 
sources. For the rest of the sample, observations are carried out at 
the Arecibo Observatory.  Integration times are set so as to 
detect HI gas mass fractions ($f_{\rm HI}=$\mhi$ / $\mstar) of 1.5\% or 
more. Observations are carried out using the $L-$band Wide receiver and 
the interim correlator, providing coverage of the full frequency 
interval of the GASS targets at a velocity resolution of 1.4 \kms \ 
before smoothing.  Data reduction includes Hanning smoothing, 
bandpass subtraction, radio frequency interference (RFI) 
excision, flux calibration and weighted combination of individual 
spectra. Total \hi-line fluxes, velocity widths and recessional 
velocities are then measured using linear fitting of the edges of 
the \hi\ profiles \citep[e.g.][]{springob05,catinella07}. 

\subsection{IRAM 30m observations}

We measure the CO(1-0) rotational transition line using the IRAM 30m 
telescope and its Eight Mixer Receiver (EMIR).  As a backend, we use 
the Wideband Line Multiple Autocorrelator (WILMA), giving a resolution of 
2 MHz ($\sim 5$ \kms\ at the observing frequency).  We also simultaneously 
record the data with the 4MHz Filterbank, as a backup.  With a single 
tuning of the receiver at a frequency of 111.4081 GHz, we are able to 
detect the redshifted CO(1-0) line for all the galaxies in our 
sample ($0.025<z<0.05$), within the 4 GHz bandwidth covered by 
the spectrometers. We integrate on each galaxy until the first of the 
following three conditions is met: (1) the CO line is detected with 
$S/N>5$, (2) sensitivity to a molecular gas mass fraction 
($M_{\rm H2}/M_{\ast}$) of 1.5\% is achieved, or (3) an absolute minimum 
rms of 1.1mK (per 20\kms-wide channel) is reached.  For most galaxies, 
the total CO line flux can be recovered by a single pointing of 
the IRAM 30m telescope. For the most extended objects, a second pointing 
is performed, offset by 16\arcsec\ along the major axis.  A set of models 
is used to apply the appropriate aperture corrections based on 
these measurements. 

The data are reduced with the CLASS/GILDAS\footnote{\tt http://www.iram.fr/IRAMFR/GILDAS} software. 
All scans are visually examined, and those with distorted baselines, 
increased noise due to poor atmospheric conditions, or anomalous 
features are discarded. The individual scans for a single galaxy are 
baseline-subtracted (first order fit) and then combined. 
This averaged spectrum is finally binned to a resolution of $\sim 20$\kms. 

The flux in the CO(1-0) line is measured by adding the signal 
within an appropriately defined windowing function.  If the line is 
detected, the window is set by hand to match the observed line profile. 
If the CO line is undetected or very weak, the window is set either to 
the full width of the HI line ($W50_{HI}$) or to a width of 
300 \kms\ in case of an HI non-detection.  The detection rate of the 
CO line is $54\%$, but our observing strategy allows us to set a 
stringent upper limit in the case of a non detection. 

After correcting the CO(1-0) line fluxes (or upper-limits) for aperture effects, using a set of models derived from a compilation of nearby galaxies with resolved maps, total CO luminosities $L^{\prime}_{CO}$ are calculated following \citet{solomon97}.  The total molecular hydrogen masses are then calculated as $M_{H2}=L'_{CO}\alpha_{CO}$.  We adopt a constant Galactic conversion factor of $\alpha_{CO}=3.2$ M$_{\odot}$(K \kms\ pc$^2)^{-1}$, which does not include a correction for the presence of Helium.  

An extensive description of 
the IRAM observing procedure and of the data reduction steps can be found in \paperI. 
The CO line fluxes and molecular gas masses for all the galaxies 
in this study are given in Table 2 of \paperI. 

\subsection{Star formation rates}

\begin{figure}
\includegraphics[width=84mm]{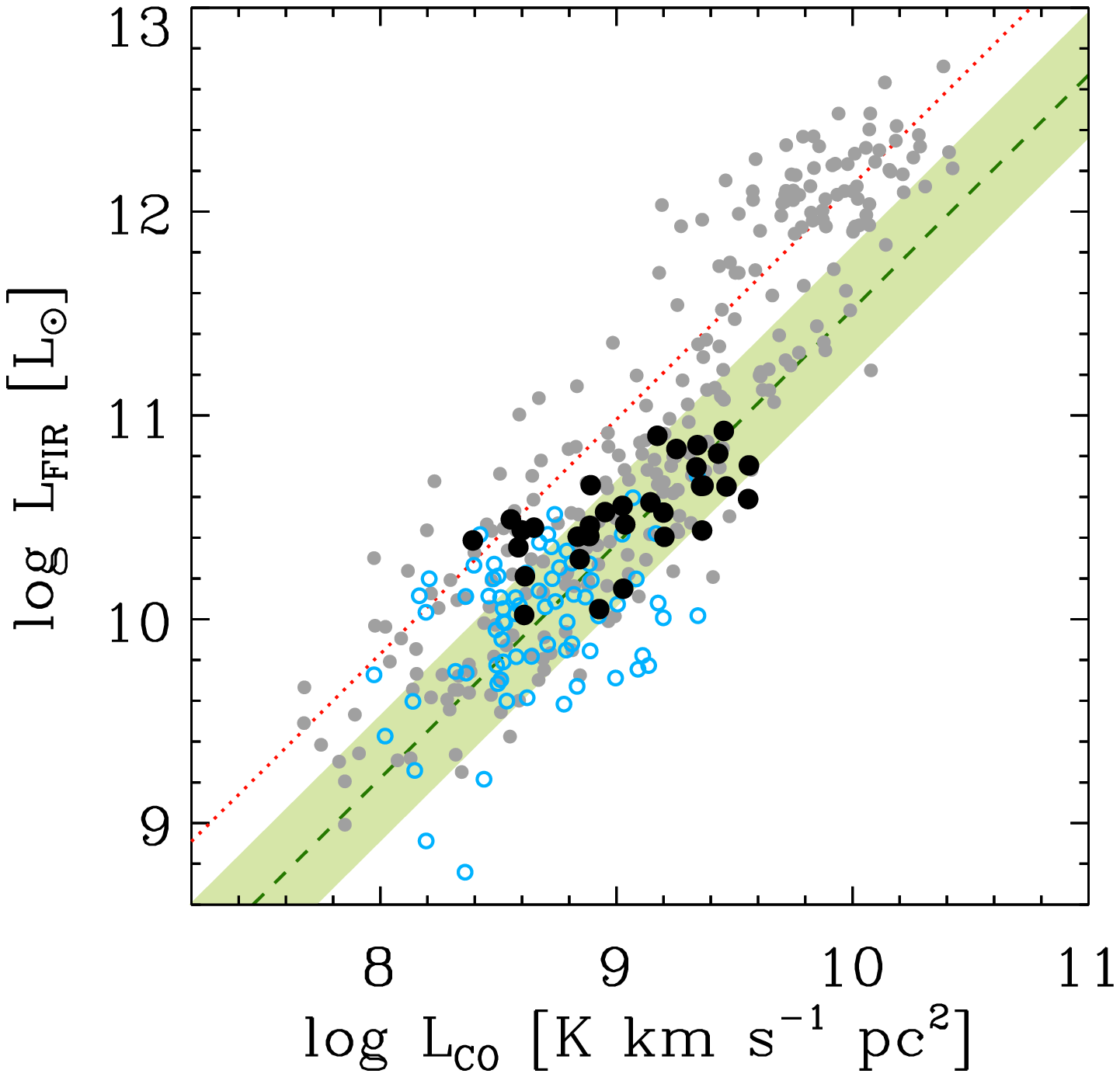}
\caption{Relation between CO luminosity and far-infrared (FIR) luminosity. 
Small gray circles show the compilation of normal and luminous 
infrared galaxies presented in \citet{gracia11}.  The dashed line 
shows the relation derived by \citet{genzel10} for normal galaxies
(the shaded region denoted the $1\sigma$  scatter about
this relation);the dotted line is the relation derived by 
the same authors for ULIRGs/merging systems.  COLD GASS galaxies with 
IRAS measurements are overplotted as filled circles and are seen to 
follow the same trend. For the remainder of the COLD GASS sample (open blue circles), 
we ``infer'' $L_{\rm FIR}$ from the total star formation rate 
after subtracting the component contributed by the observed UV
radiation (without attenuation correction).  
  \label{LIRLCO}}
\end{figure}

Star formation rates for all COLD GASS galaxies are calculated from 
optical/UV spectral energy distribution (SED) fitting (\sfrsed).  
The technique is explained in detail in Appendix \ref{SFRap}.   In short, 
fluxes in the seven SDSS and GALEX bands are fitted with a series of 
model SEDs constructed from the \citet{bc03} population synthesis code, 
assuming a range of star formation histories, ages, metallicities
and dust attenuation factors. As described in the Appendix, the
assumed distribution of dust attenuation factors is the critical factor
in obtaining accurate star formation rates using this method. We use the 
sample of \citet{johnson07} to calibrate our dust attenuation
priors. This dataset includes GALEX FUV data, 
plus 8, 24 and 70 $\mu$m data from Spitzer for a representative
sample of galaxies selected from SDSS.  
For galaxies with \nuvr$<5$, our values of \sfrsed\ are consistent 
with the total star formation rates (\sfrtot) derived from the combined 
GALEX FUV and Spitzer 70$\mu$m fluxes, with scatter of 
only 0.22 dex (see Appendix \ref{SFRap} and Figure \ref{compsfr}). 

\subsection{Comparison with infrared luminosity}

For a small fraction of the COLD GASS sample ($\sim 15\%$), 
flux measurements at 60 and 100 $\mu$m can be found in the IRAS Faint 
Source Catalog \citep{moshir92}.  The far-infrared luminosity inferred 
from these fluxes is an excellent measurement of the total luminosity
from young stars that has been reprocessed by dust, 
and is known to correlate well with the total CO luminosity of the 
galaxies \citep[e.g.][]{sanders85,sanders91,gao04}. 
In Figure \ref{LIRLCO}, we plot the $L_{FIR}-L_{CO}$ relation for 
the COLD GASS sample and compare it to literature results for a collection 
of nearby galaxies.  The COLD GASS galaxies with IRAS measurements 
(filled circles in Figure \ref{LIRLCO}) follow previously derived
relations almost exactly. The rms scatter about the relation is 0.30 dex, 
compared to the value of 0.31 dex derived for reference samples 
of nearby galaxies \citep{genzel10}. 

For the COLD GASS galaxies without IRAS measurements, 
we {\em infer} $L_{FIR}$ by assuming that it should
be equal to (SFR$_{SED}$-SFR$_{UV}$).  These estimates are
plotted as  open symbols in Figure \ref{LIRLCO}. For the sample,
the scatter around the \citet{genzel10} relation is 0.39 dex. This enhanced 
scatter could be due to the uncertainty in the values of 
\sfrsed (see Appendix \ref{SFRap});  only direct infrared 
measurements of these galaxies will allow us to firmly 
establish the level of scatter in the $L_{CO}-L_{FIR}$ relation 
for our complete sample. Nevertheless, Figure \ref{LIRLCO}
makes it clear that our sample spans a broader range in properties
such as  star formation rate compared to IRAS-selected samples.
As we will show in Section \ref{results}, our samples also include  galaxies with
a wider range of CO-to-IR luminosities or, equivalently, molecular gas mass 
to star formation rate ratios.    

\subsection{Molecular gas depletion timescales}

Using the molecular hydrogen gas masses (or upper limits) and the star formation rates derived from SED fitting, we define for each galaxy its molecular gas depletion timescale:
\begin{equation}
t_{dep}({\rm H_2}) \equiv M_{\rm H_2} {\rm SFR_{SED}}^{-1}.
\label{depeq}
\end{equation}
This parameter therefore describes how long each galaxy could sustain 
star formation at the current rate before running out of fuel, 
assuming that the gas reservoir is not replenished. Some studies instead 
consider the inverse of \tdep, the molecular gas star formation 
efficiency (SFE$_{H_2}$).  However, since the definition of SFE is not 
universal, with some authors quoting SFE as a dimensionless quantity 
per dynamical time, we use in this study the more straightforwardly-defined 
depletion timescale, defined in Equation \ref{depeq}. 
Similarly, the atomic gas depletion timescale is 
\tdepHI$\equiv M_{\rm HI} {\rm SFR_{SED}}^{-1}$.  

\subsection{Sample definition}

\begin{figure}
\includegraphics[width=84mm]{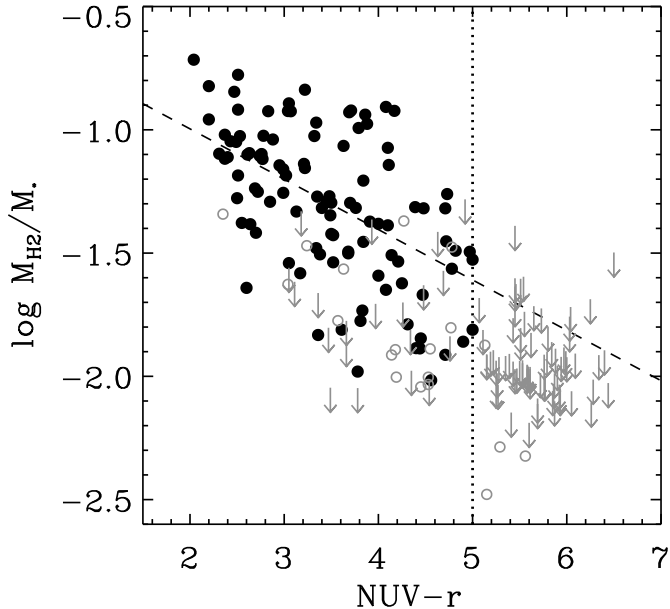}
\caption{Molecular gas mass fraction as a function of \nuvr\ colour for the current COLD GASS sample of \ntot\ galaxies. Secure detections are represented as filled circles, tentative detections as open circles, and non-detections as $5\sigma$ upper limits as indicated by the arrows.  The best fit linear regression is overplotted, as well as the empirical threshold of \nuvr$=5$ which marks an abrupt transition from the gas-rich, star forming population to a more passive population.  
\label{nuvr}}
\end{figure}

One of the key results of \paperI\ is presented in Figure \ref{nuvr}.  We found sharp thresholds in galaxy parameters such as stellar mass (\mstar), stellar mass surface density (\must), concentration index ($R_{90}/R_{50}$) and colour, below which almost all galaxies have a measurable cold gas component but above which the detection rate of the CO line drops suddenly.  These thresholds correspond approximately to the transition between blue cloud and red sequence.  The strongest bimodality is seen when relating the molecular gas mass fraction and the \nuvr\ colour of the galaxies (Figure \ref{nuvr}).  A clear threshold at \nuvr=5 marks a break between the star forming and the passive populations.   

This colour threshold is also the limit where the uncertainty on the SED fitting-derived star formation rates increases abruptly (Figure \ref{compsfr}).  Galaxies with \nuvr$>5$ only have upper limits for \mh\ and very uncertain measures of SFR, therefore their depletion times (Equation \ref{depeq}) are very poorly constrained. From here on, we therefore consider only the subset of COLD GASS galaxies which have \nuvr$<5$ and which make up the active population.  Through stacking, we have shown in \paperI\ that the passive population is truly ``red and dead", with a stringent constraint on their average gas mass fraction of $M_{H_2}/M_{\ast}=0.0016\pm0.0005$. These red galaxies are therefore mostly irrelevant for the following discussion and we do not consider them further.  Galaxies with \nuvr$<5$ but no CO detection are considered as upper limits in the \tdep\ plots, using $5\sigma$ upper limits for \mh\ (see \paperI).

\section{A non-universal molecular gas depletion timescale}
\label{results}

Resolved studies of the disks of normal galaxies indicate that 
the molecular gas depletion timescale is constant \citep{bigiel08,leroy08}. 
In this section, we investigate how our integrated measurements 
of \tdep\ scale with a number of other global physical parameters,
including stellar mass, stellar surface mass surface density, 
bulge-to-disk ratio and specific star formation rate.

\subsection{Variations with global physical parameters}
\label{tdeprelations}

\begin{figure*}
\begin{minipage}{165mm}
\includegraphics[width=165mm]{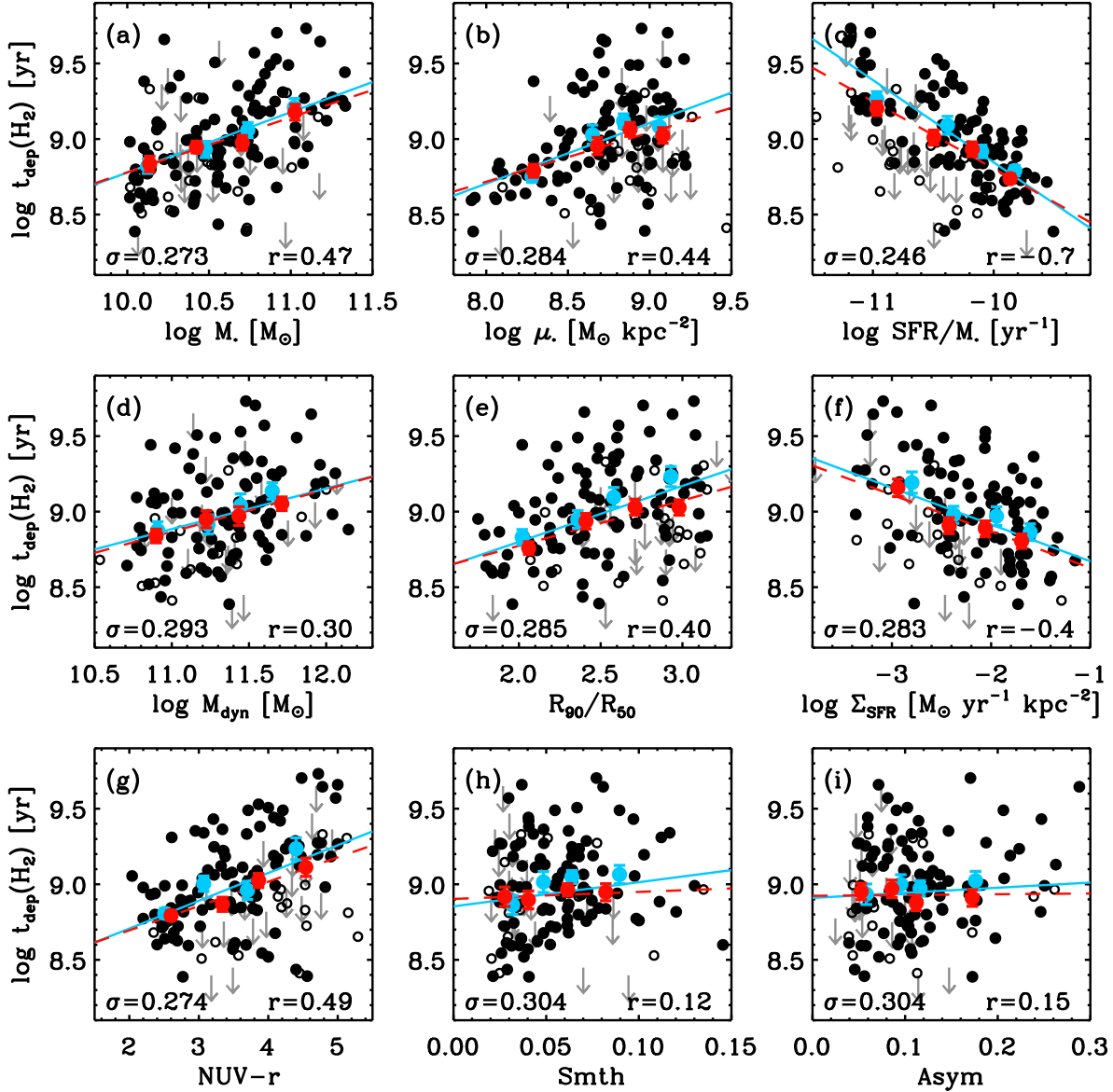}
\caption{Depletion time for the molecular gas as a function 
of a variety of  parameters. Filled and open black circles indicate
COLD GASS secure and tentative detections. 
We also show the mean values in equally populated bins for secure detections only (blue symbols) 
and for the entire sample with \nuvr$<5$ (red symbols). 
The error bars indicate 1$\sigma$  confidence interval on the mean, 
determined by bootstraping.  The solid and dashed lines are the linear regression 
of \tdep\ against the $x-$axis parameter, for the 
secure detections only and the entire sample, respectively.  In each panel we give the scatter in the residuals 
of this best-fit relation ($\sigma$), as well as the 
correlation coefficient measured for the secure detections ($r$).
 \label{depall}}
 \end{minipage}
\end{figure*}

In Figure \ref{depall}, the molecular gas depletion timescale is 
plotted against a range of parameters describing the global physical
properties of the galaxies.  Although there is scatter, a clear 
dependence of \tdep\ on many of the parameters is observed.  
In all cases, we use two different measures to quantify the strength 
of the dependence of \tdep\ on the x-axis parameter: (1) the Pearson 
correlation coefficient of the parameters, $r$, and (2) the scatter of 
the residuals in $\log($\tdep) of the best fit linear relation (regression 
of $y$ against $x$), $\sigma$.  As a comparison point, the scatter 
in $\log($\tdep) for all the CO detections is $\sigma=0.315$ dex.  We fit both to the secure detections only, and also to the entire sample of galaxies with \nuvr$<5$ including tentative detections and upper limits. 
 The coefficients as well as the parameters of the best-fit 
relations for both of these (sub-)samples are given in Table \ref{tdeptab}.   All the fits and mean values are computed by weighting galaxies differently according to their stellar mass.  This procedure, explained in detail in \paperI, corrects for the fact that by selection the COLD GASS sample has a $\log M_{\ast}$ distribution which is flatter than in a purely volume-limited sample. 

\begin{table*}
\begin{minipage}{165mm}
\caption{Molecular gas depletion time relations$^a$}
\label{tdeptab}
{\scriptsize
\begin{tabular}{cccccccccccc}
\hline
 & & & \multicolumn{4}{c}{All galaxies with NUV$-r<5$} & & \multicolumn{4}{c}{Secure detections only} \\
\cline{4-7} \cline{9-12} \noalign{\smallskip}
$x$ parameter & units & $x_0$ & $m$ & $b$ &
$\sigma$ & $r$  &  & $m$ & $b$ & $\sigma$ & $r$ \\
\hline
                    $\log M_{\ast}$ &                              $\log M_{\odot}$ &  10.70 & $  0.36  \pm   0.07$ & $  9.03  \pm   0.99$ &  0.274 &   0.42 &
       & $  0.40  \pm   0.07$ & $  9.06  \pm   1.03$ &  0.269 &   0.48 \\
                  $\log \mu_{\ast}$ &               $\log M_{\odot} {\rm kpc}^{-2}$ &   8.70 & $  0.33  \pm   0.07$ & $  8.95  \pm   0.91$ &  0.284 &   0.36 &
       & $  0.40  \pm   0.07$ & $  8.99  \pm   0.88$ &  0.275 &   0.44 \\
          $\log {\rm SFR}/M_{\ast}$ &                          $\log {\rm yr}^{-1}$ & -10.40 & $ -0.44  \pm   0.04$ & $  8.98  \pm   0.61$ &  0.246 &  -0.61 &
       & $ -0.54  \pm   0.04$ & $  9.06  \pm   0.62$ &  0.221 &  -0.70 \\
                     $\log M_{dyn}$ &                              $\log M_{\odot}$ &  11.40 & $  0.28  \pm   0.07$ & $  8.98  \pm   1.05$ &  0.293 &   0.31 &
       & $  0.27  \pm   0.08$ & $  8.99  \pm   1.16$ &  0.291 &   0.31 \\
                  R$_{90}/$R$_{50}$ &                                       \nodata &   2.50 & $  0.30  \pm   0.06$ & $  8.92  \pm   0.21$ &  0.286 &   0.34 &
       & $  0.37  \pm   0.07$ & $  8.99  \pm   0.22$ &  0.290 &   0.40 \\
            $\log \Sigma_{\rm SFR}$ & $\log M_{\odot} {\rm yr}^{-1} {\rm kpc}^{-2}$ &  -2.40 & $ -0.24  \pm   0.04$ & $  8.97  \pm   0.15$ &  0.283 &  -0.43 &
       & $ -0.24  \pm   0.06$ & $  9.01  \pm   0.16$ &  0.287 &  -0.44 \\
                            NUV$-r$ &                                           mag &   3.50 & $  0.16  \pm   0.03$ & $  8.94  \pm   0.14$ &  0.275 &   0.42 &
       & $  0.18  \pm   0.03$ & $  8.98  \pm   0.15$ &  0.263 &   0.49 \\
                               Smth &                                       \nodata &   0.05 & $  0.46  \pm   1.01$ & $  8.93  \pm   0.08$ &  0.304 &   0.06 &
       & $  1.60  \pm   1.28$ & $  8.93  \pm   0.10$ &  0.307 &   0.13 \\
                              Asym &                                       \nodata &   0.10 & $  0.05  \pm   0.49$ & $  8.93  \pm   0.08$ &  0.305 &   0.10 &
       & $  0.34  \pm   0.61$ & $  8.94  \pm   0.09$ &  0.309 &   0.16 \\
\hline
\end{tabular}
}
$^a$ The relations are parametrized as $\log$ \tdep[yr$^{-1}]=m(x-x_0)+b$, with all quantities and units given in this table.
\end{minipage}
\end{table*}

Interestingly, the depletion timescale depends both on parameters that 
describe the masses and structural properties of the galaxies and on parameters 
relating to their stellar populations and star formation rates. 
 A clear dependence of \tdep\ on the stellar mass of the systems is 
observed ($r=0.48$), as well as on the presence of a bulge as 
parametrized by the concentration index, $R_{90}/R_{50}$ ($r=0.40$), 
and on pressure within the disk as measured by the stellar mass surface 
density, \must\ ($r=0.44$).  In all these cases, the scatter in the 
residuals of the best-fit relations is significantly smaller 
than the scatter in \tdep\ itself.  Of these three, the dependence on $\log M_{\ast}$ is strongest, and also the most insensitive to the presence of non-detections in CO (see Figure \ref{depall}a and Table \ref{tdeptab}). 

There is a weaker dependence of \tdep\ on dynamical mass 
(Figure \ref{depall}d, $r=0.31$), which is estimated as: 
\begin{equation}
M_{dyn}=\frac{v_c^2 R}{G},
\label{mdyn}
\end{equation}
where the circular velocities, $v_c$, are measured from the \hi\ linewidths, 
$W50_{HI}$, as:
\begin{equation}
v_c=\frac{W50_{HI}/2.0}{(1+z) \sin({\rm incl})}.
\end{equation}
The factor $(1+z)$ accounts for cosmological stretching 
and the $\sin({\rm incl})$ term is a correction for the inclination 
of the system (as measured from SDSS photometry).  
In Equation \ref{mdyn}, we further assume that the size of the 
\hi\ disk ($R$) is 1.5 times the size of the optical disk ($D_{25}/2$). 
Uncertainty in this approximation certainly contributes 
to the scatter of the relation between \tdep\ and $M_{dyn}$.  

The strongest dependencies in Figure \ref{depall} are however on
quantities that are sensitive to the amount of
star formation in the galaxies.  The two strongest correlations 
of \tdep\ are with \nuvr\ colour ($r=0.49$) and with specific star 
formation rate (sSFR, $r=-0.7$).  These quantities are directly linked, 
because \nuvr\ is a good proxy for sSFR since it relates a 
quantity tracing ongoing star formation activity (NUV flux) 
and a quantity sensitive to the older stellar population 
($r-$band flux).  However, they differ in the sense 
that the sSFR takes into account internal dust attenuation and
the mass-to-light ratio of the old stellar population, 
while \nuvr\ is not corrected for both these effects.  

\begin{figure*}
\begin{minipage}{165mm}
\includegraphics[width=165mm]{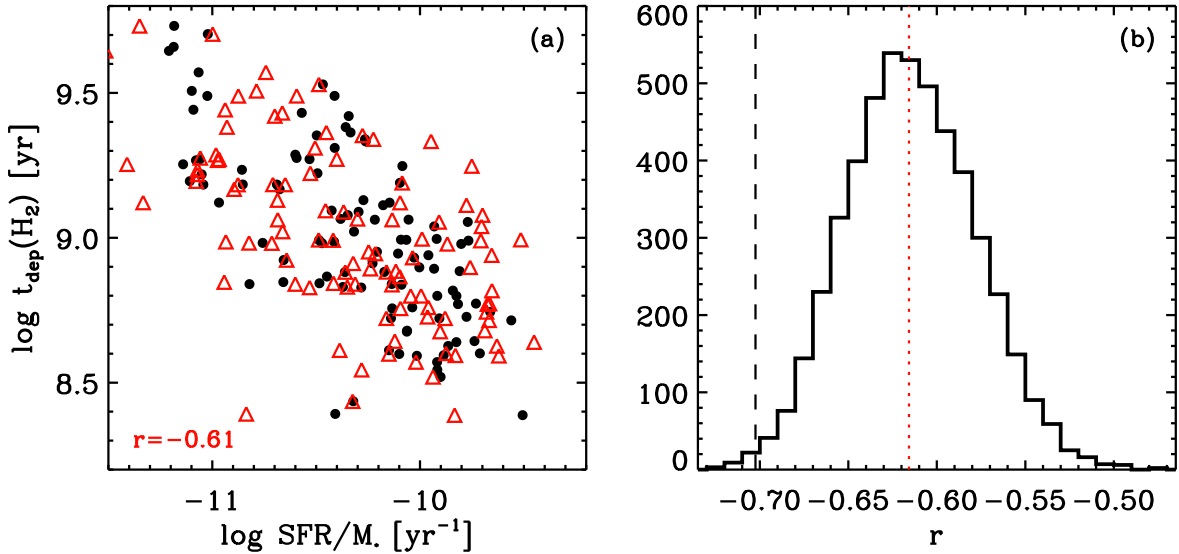}
\caption{Validation of the \tdep-sSFR relation using a Monte-Carlo approach.  The correlation coefficient ($r$) of the original relation is compared to the value measured for a similar relation after decorrelation of the measurement errors. {\bf (a)} An example showing one of the repetitions and comparing the original relation (filled circles) with the ``decorrelated" relation (open triangles).  {\bf (b)} Distribution of the correlation coefficient of the 5000 iterations (median $r=-0.615$).  While there is only a $0.5\%$ chance that the original strength of the correlation ($r=-0.71$) is not due to measurement correlation, the probability that the relation is  more strongly correlated than the \tdep-\nuvr\ or the \tdep-$M_{\ast}$ relations is 99.9\%.
 \label{montecarlo}}
 \end{minipage}
\end{figure*}

At first glance, \tdep\ seems to depend more strongly on sSFR than on \nuvr\ colour, as revealed by the correlation coefficients and scatter around the best fit relations (Table \ref{tdeptab}).  However, we must be cautious because the same measure of the SFR enters in the computation of \tdep\ and sSFR, and the stronger correlation with sSFR compared to \nuvr\ may therefore only be induced by this common quantity.  We test for this using a Monte-Carlo approach, which aims to decorrelate the errors on \tdep\ and sSFR.  To mimic an independent measure of the two parameters, while keeping \tdep\ fixed to the original values, we recompute the sSFR by allowing the star formation rate of each galaxy to vary around the measured value according to the observed measurement error, assuming a normal distribution ($\sigma_{SFR}=0.24$ dex, see Figure \ref{compsfr}).   We perform 5000 iterations and examine the distribution of the correlation coefficients in the simulated relations, which has a median value of $r=-0.61$ as shown in Figure \ref{montecarlo}.  While there is only a $0.5\%$ chance that the original strength of the correlation ($r=-0.71$) is not due to measurement correlation, the probability that the relation is  more strongly correlated than the \tdep-\nuvr\ ($r=0.49$) or the \tdep-$M_{\ast}$ ($r=0.48$) relations is $>99.9$\%.  We therefore conclude that while part of the strength of the correlation between \tdep\ and sSFR is indeed induced by the common measurement of SFR entering both parameters, the relation is still at least if not more strongly correlated than the next most significant correlations (i.e. those on \nuvr\ colour and stellar mass).  In Section \ref{specvalid} below, we present further evidence for this. 

The molecular gas depletion time is also correlated with the star 
formation surface density, measured within $R_{50}$(SFR),
the radius enclosing half of the total 
star formation measured for the whole galaxy. This
quantity is measured as part of the \sfrsed\ calculation pipeline. 
The gas surface density, $\Sigma_{gas}$, cannot be accurately 
calculated for the COLD GASS sample since no direct measurement 
of the size of the gas disks is available.  
However, since $\Sigma_{gas}$ and $\Sigma_{SFR}$ are 
directly linked through the Kennicutt-Schmidt relation, we can 
infer from Figure \ref{depall}f that \tdep\ is a decreasing 
function of $\Sigma_{gas}$. 

On the other hand, \tdep\ does not appear to depend on measures 
of the degree of asymmetry and clumpiness in the galaxies 
(Figure \ref{depall}h,i), the scatter of the residuals of these 
two relations ($\sigma=$0.308 and 0.306, respectively) is not 
significantly reduced compared to the scatter of 0.315 dex in \tdep\
for the whole sample. Such dependencies might have been expected if, 
for example, spiral density waves or other perturbations in the disk
trigger the formation of GMCs and thus 
contribute locally to a decrease in \tdep.  
The fact that we do not see a clear dependene of \tdep\ 
on our measurements of structure within the disks 
could mean either that $g-$band light (the band used to measure Asym 
and Smth) is not an optimal tracer of these compression events, 
or else that the effect mostly disappears once
averaged over entire galaxies.  We note however that this independence of \tdep\ on spiral arm patterns was also reported by \citet{leroy08} using resolved data.

\begin{figure*}
\begin{minipage}{165mm}
\begin{center}
\includegraphics[width=125mm]{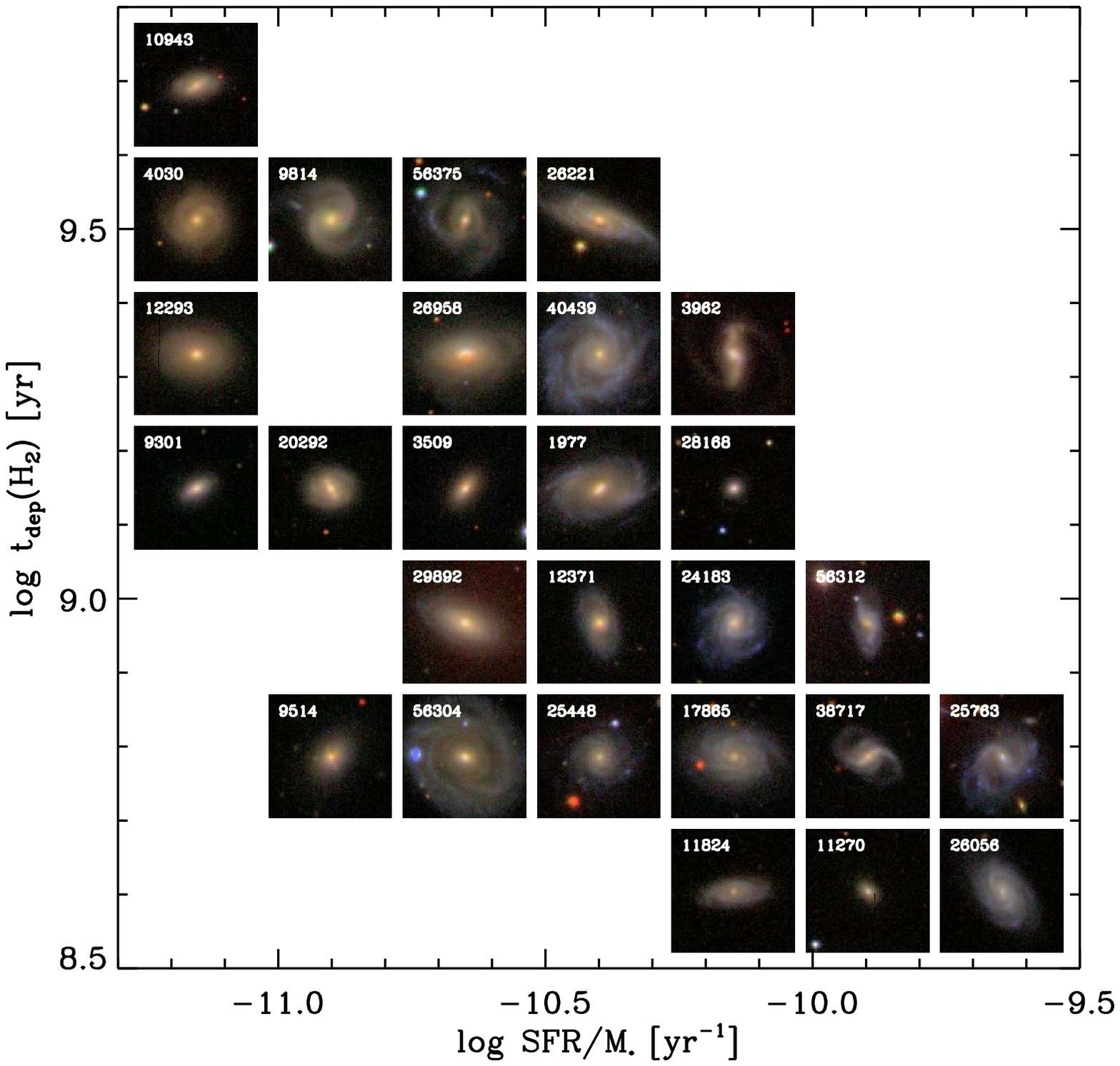}
\caption{SDSS three-colour images (of dimension 50$\times$50 kpc) of
a subset of the COLD GASS galaxies in the sSFR-\tdep\ plane. 
 Some of the trends present in Figure \ref{depall} are visible:  
galaxies with the longest depletion times tend to be redder,  
with early-type morphologies and more prominent bulges. 
 \label{gallery}}
 \end{center}
 \end{minipage}
\end{figure*}

To illustrate some of the results presented in 
Figure \ref{depall} graphically,
we show in Figure \ref{gallery} a gallery of SDSS thumbnails 
of COLD GASS galaxies at their appropriate locations in the 
sSFR-\tdep\ plane. The images clearly reveal that 
galaxies with longer molecular gas depletion times tend 
to be redder, more massive, and have larger bulges than those with shorter depletion
times. 

As a final step, we investigate whether third parameter dependences 
can be found. We analyze whether the residuals of the sSFR-\tdep\ 
relation (the tightest correlation found in Figure \ref{depall}) show
any dependence on any of the parameters included in 
this study (see Figure \ref{residuals} and details in Appendix \ref{otherparams}).  
We do not identify any convincing third parameter dependence.

Our results imply that for a given amount of molecular gas, all galaxies 
are not equally efficient at processing this gas into stars. 
We find an  increase 
in the mean molecular gas depletion timescale by a factor of 
$\sim 6$ over the stellar mass range of $10^{10}$ to $10^{11.5}$\msun.  
This implies that galaxies with large stellar  masses, red colours, high stellar 
surface densities and/or older stellar populations will require 
significantly longer times to turn their molecular gas into stars.  
The mean depletion timescale in these systems is \tdep$\sim 3$ Gyr, 
as compared to the average of $\sim1$ Gyr for the sample as a whole,
and to the mean value of $\sim 0.4$ Gyr for galaxies with
stellar masses $\sim 10^{10}$\msun.

\subsection{Validation using spectroscopic parameters}
\label{specvalid}

\begin{figure*}
\begin{minipage}{165mm}
\includegraphics[width=165mm]{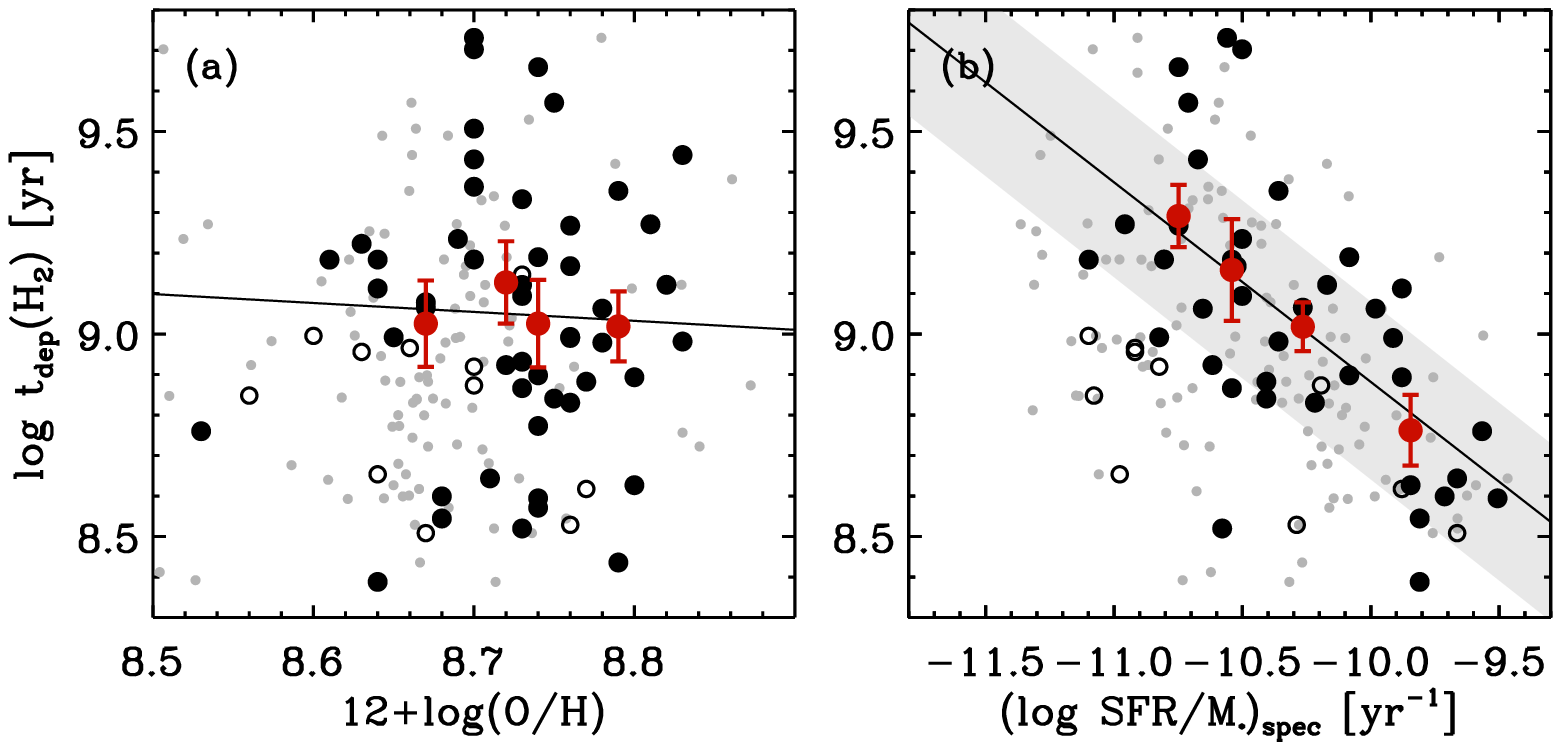}
\caption{Molecular gas depletion time as a function of two 
spectroscopic quantities: (a) the gas metallicity calculated using 
Equation \ref{Zeq} and (b) the sSFR inferred from the measurement 
of the 4000\AA\ break (see text). The spectroscopic quantities are derived 
from the SDSS fiber spectroscopy (small gray symbols) and 
independently from the long-slit spectra (large black symbols, 
open for tentative CO detections).  The solid lines are the 
best-fitting relations to the latter.  In panel (b), as a comparison, 
the 1$\sigma$ scatter around the relation of Figure \ref{depall}c derived using  
sSFR$_{SED}$ is shown as the shaded region. 
The spectroscopic data provide an independent confirmation 
of the trend observed between sSFR and \tdep. 
\label{depspec}}
\end{minipage}
\end{figure*}                            

In this section, we analyze the behavior of \tdep\ as a function of
two  spectroscopically-derived quantities, which we measure either 
from SDSS fiber spectroscopy or from the long slit data 
(see Section \ref{spectra}).  While SDSS fiber spectra are available 
for the entire COLD GASS sample, they have the disadvantage of 
sampling only the central 3\arcsec of the galaxies.  This 
limitation could bias our results when investigating quantities 
that vary radially within the disks. 

We measure metallicity from strong line ratios using the
prescription of \citet{pettini04}:
\begin{equation}
12+\log ({\rm O/H}) = 8.73-0.32\log \left ( \frac{{\rm [OIII]/H}_{\beta}}{{\rm [NII]/H}_{\alpha}} \right ).
\label{Zeq}
\end{equation}
On this scale, a value of $12+\log ({\rm O/H})=8.66$ corresponds to solar metallicity. 
In Figure \ref{depspec}a, we plot \tdep\ as a function of metallicity
measured from both sets of spectra. 
\tdep\ is found to be completely independent of metallicity 
(the correlation coefficient is $r=0.01$). In fact, of 
all parameters investigated (e.g. in Figure \ref{depall}),
metallicity is the one that shows the least degree of correlation 
with the depletion timescale.  This result is reassuring, 
as a metallicity dependence could have cast some doubt 
on our assumption of a constant CO-to-\hmol\ 
conversion factor ($X_{CO}$). We remind the reader that the
COLD GASS sample only includes galaxies with stellar masses greater
than $10^{10} M_{\odot}$ and less than
a few $\times 10^{11} M_{\odot}$. The relation between metallicity and stellar mass   
is rather flat over this regime (Tremonti et al 2004), with the average
gas-phase metallicity only increasing by $\sim 0.25$ dex over 1.5 dex in $\log M_*$. 

The spectroscopic data also provide us with information
about stellar absorption features, including the strength of the 
4000\AA\ break. The strength of this feature varies montotonically with
the age of a stellar population of fixed metallicity. If we assume
that the galaxies in our sample have had smooth, exponentially declining
star formation histories, we can use standard stellar population
synthesis models (Bruzual \& Charlot 2003) to calibrate a
relation between the 4000 \AA\ break strength and sSFR. 
We have done this both for the  fiber and the long-slit measurements
of this feature.   

This independent estimate of the sSFR can be used as a second validation of 
the strength of the dependence of \tdep\ on the 
specific star formation rate.  In Figure \ref{depspec}b, we show that 
the relation between \tdep\ and sSFR estimated from
the 4000 \AA\ break has a correlation coeffcient  of -0.61,
almost as strong as the value of -0.70 measured for the same 
relation using photometric quantities, and identical to the value estimated using the Monte-Carlo simulations (see Figure \ref{montecarlo}b). 
This spectroscopic measurement therefore provides independent 
confirmation of the effects already seen with 
\nuvr\ colour (Figure \ref{depall}g) and photometrically-derived 
sSFR (Figure \ref{depall}c).

\subsection{Comparison with resolved studies}

\begin{figure}
\includegraphics[width=84mm]{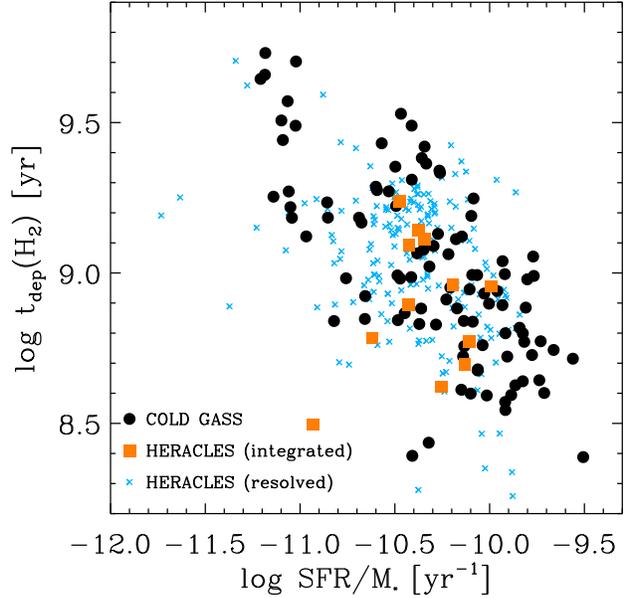}
\caption{Depletion time for the molecular gas as a function of specific 
star formation rate for COLD GASS secure  
detections (filled circles) as well as 
for the HERACLES sample from \citet{leroy08}.  We plot both the integrated data (filled squares) and the resolved data (crosses).  The COLD GASS sample spans a larger range in global values of sSFR than the HERACLES sample.  When considering integrated measurements, both data sets however show similar scatter and behaviour in the range of sSFR where they overlap. \label{compHERA}}
\end{figure}

We compare our results with those obtained from resolved studies of nearby galaxies.  The two approaches are highly complementary: COLD GASS gives us a global picture of gas and star formation in the local Universe, but lacks the power to trace the exact distribution of these components, something surveys like THINGS/HERACLES are designed to do.  The methods are however fundamentally distinct, since a volume-limited survey like COLD GASS gives more statistical weight to the lower mass galaxies which naturally dominate the galaxy number density, while a resolved study tends to give more weight to bigger, more massive galaxies which cover more resolution elements.  These distinctions should be kept in mind when comparing observations.

While our results show a variation in \tdep, in contrast with the constant \hmol\ depletion time found for HERACLES survey galaxies \citep{leroy08,bigiel08}, the results are not surprising. HERACLES covers a limited range of surface densities typical of nearby ``normal" spirals. Observations that probe a wider range of galaxy types have long been known to give different H$_2$ depletion times between the regimes of normal  and  ``bursty" star formation \citep[e.g. in mergers and ULIRGS,][]{gao04}, and theoretical models have been able to explain this dichotomy as arising from a transition between galaxies where molecular clouds are gravity-confined to ones where they are pressure-confined \citep{krumholz09}. In order to illustrate the consistency of our results with HERACLES, in Figure \ref{compHERA} we directly 
compare the two data sets.  To do this comparison, 
we plot the depletion timescale as a function of 
specific star formation rate (sSFR), since this quantity can be 
reliably compared between the two samples.  The comparison shows that the  
COLD GASS data are perfectly consistent with the 
HERACLES measurements, once all quantities are brought to 
a common scale (same conversion factor $\alpha_{CO}=3.2$\msun\ (K \kms\ pc$^2$)$^{-1}$).   
The reasons why COLD GASS reveals a variable molecular gas depletion 
timescale are (1) a significantly larger sample, and (2) much larger dynamic 
range in quantities such as sSFR.   We note that \citet{leroy08} 
quote a mean molecular gas depletion time of $1.9\pm0.9$ Gyr. However, after using the same conversion factor and restricting their sample to the COLD GASS stellar mass range, the mean molecular gas depletion time for the HERACLES sample is 0.9 Gyr, perfectly consistent with the COLD GASS estimate ($1.05\pm0.74$ Gyr), as seen in Figure \ref{compHERA}.

We can gain some further insight by looking at the resolved data from HERACLES, also plotted in Figure \ref{compHERA}.  Not surprisingly, there is larger scatter in the \tdep-sSFR relation, since these points are averages over only $\sim$kpc-sized regions and not entire galaxies.   The resolved data do cover the full range of \tdep\ and sSFR values probed by the COLD GASS galaxies, but the vast majority of the data points are, however, concentrated in the range (-11.0$<\log {\rm SFR/M}_{\ast}<-10.0$) and ($8.7<\log~t_{dep}({\rm H_2}) < 9.3$).  There are very few regions within these normal star forming spirals that reach values of \tdep\ as low or as high as those seen in the {\em integrated} COLD GASS measurements. This suggests that large variations of \tdep\ between different kpc-sized regions of the same galaxies are not common \citep[see also][]{leroy08}, and therefore that extreme values on small scales are almost exclusively found in galaxies that are also globally extreme.  The few resolved points with depletion times that are unusually high (low) are located in the disk centers (outskirts), where the $M_{H2}/M_{HI}$ ratio is high (low).

\subsection{Comparison with atomic gas depletion time}

Another important step in understanding cold gas depletion times in
galaxies is the atomic-to-molecular transition. 
In this section, we compare the molecular and atomic gas 
depletion timescales across the COLD GASS sample. \citet{GASS2} studied 
the behavior of \tdepHI\ for galaxies in the GASS survey. 
They found that the average atomic gas depletion timescale for galaxies
with $M_{\ast}> 10^{10} M_{\odot}$ is 3 Gyr. There is considerable
scatter from one galaxy to another, but the average timescale is  
a constant function of all quantities investigated.  

\begin{figure*}
\begin{minipage}{165mm}
\includegraphics[width=165mm]{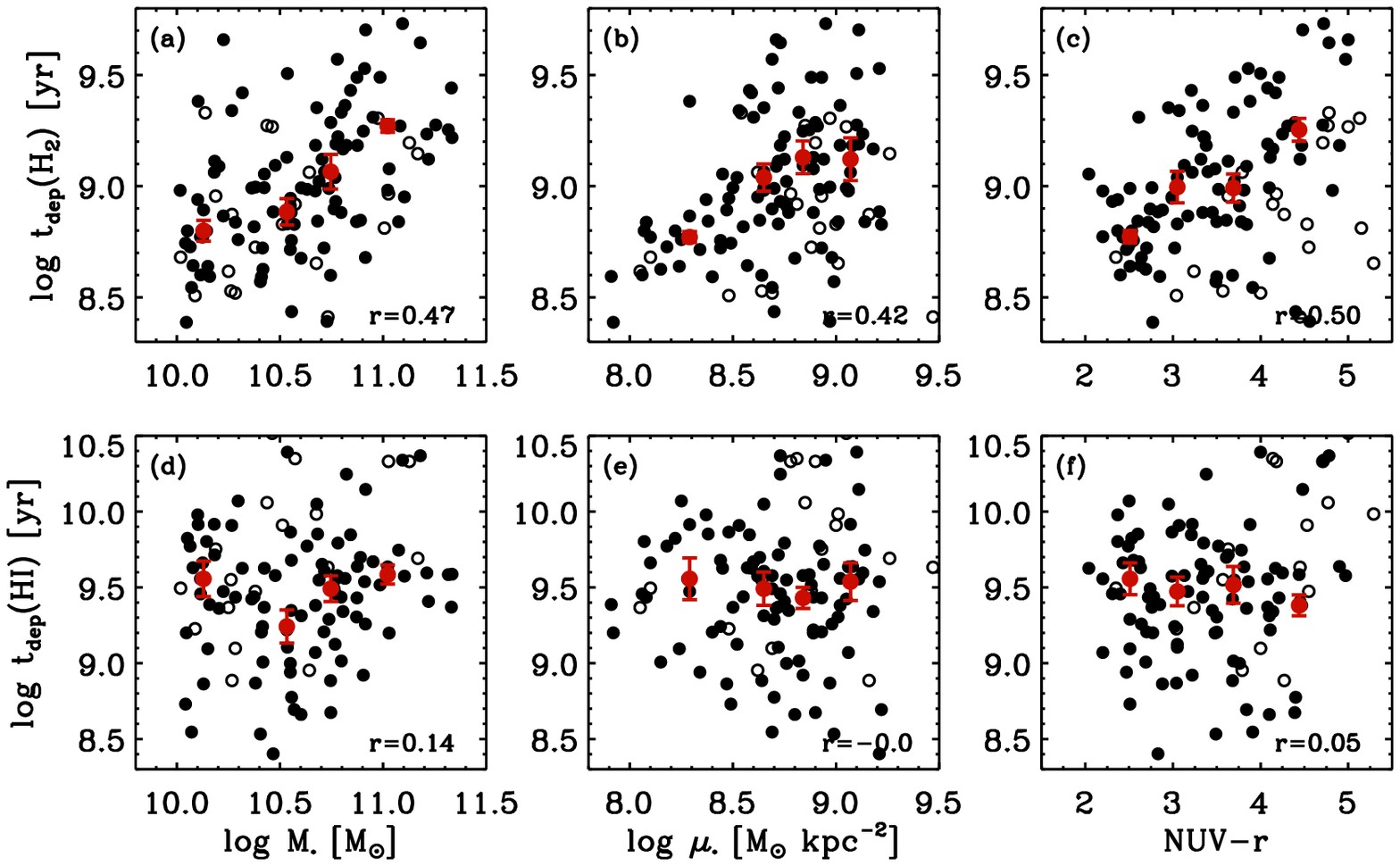}
\caption{Depletion time for the molecular gas (top row) and atomic 
gas (bottom row), as a function of {\it (a,d)} stellar mass, 
{\it (b,e)} stellar mass surface density and {\it (c,f)} NUV$-r$ colour.  \label{depHIH2}}
\end{minipage}
\end{figure*}

In Figure \ref{depHIH2} we compare the dependencies of 
\tdep\ and \tdepHI\ on \mstar, \must\ and \nuvr\ colour.  Not surprisingly, 
we reproduce well the \citet{GASS2} result; \tdepHI\ does not 
depend on any of these three quantities and is on average 3 Gyr, 
although the values of \tdepHI\ for individual galaxies range from 
a few hundred million years to more than a Hubble time.  
 In comparison, the range of values for \tdep\ is smaller, 
from a few hundred million years to $\sim 5$ Gyr. 
 This is not surprising, given that the atomic gas-to-stellar mass 
ratio has much larger dynamic range than the molecular 
gas-to-stellar mass ratio \citep[][\paperI]{GASS1}.  

Using the exact same sample, we can see clearly in Figure \ref{depHIH2} 
that whereas \tdepHI\ is constant at $\sim 3$ Gyr, \tdep\ is a 
function of all three quantities (\mstar, \must\ and \nuvr). 
It is interesting that at the highest masses, largest surface densities 
and reddest colours, \tdep$\simeq$\tdepHI\ for the majority of
galaxies.  However, for low mass, low stellar surface
density and high sSFR galaxies, \tdep\ is smaller than \tdepHI\ 
by almost an order of magnitude.  In other words, low mass galaxies
are much more imminently in danger of running out of molecular gas
than out of atomic gas, whereas in high mass systems  
there appears to be a balance between the consumption rate of atomic 
and molecular gas (at least when averaged over entire galaxies).

\section{Bridging low and high redshift molecular gas studies}
\label{highz}

\begin{figure*}
\begin{minipage}{165mm}
\begin{center}
\includegraphics[width=135mm]{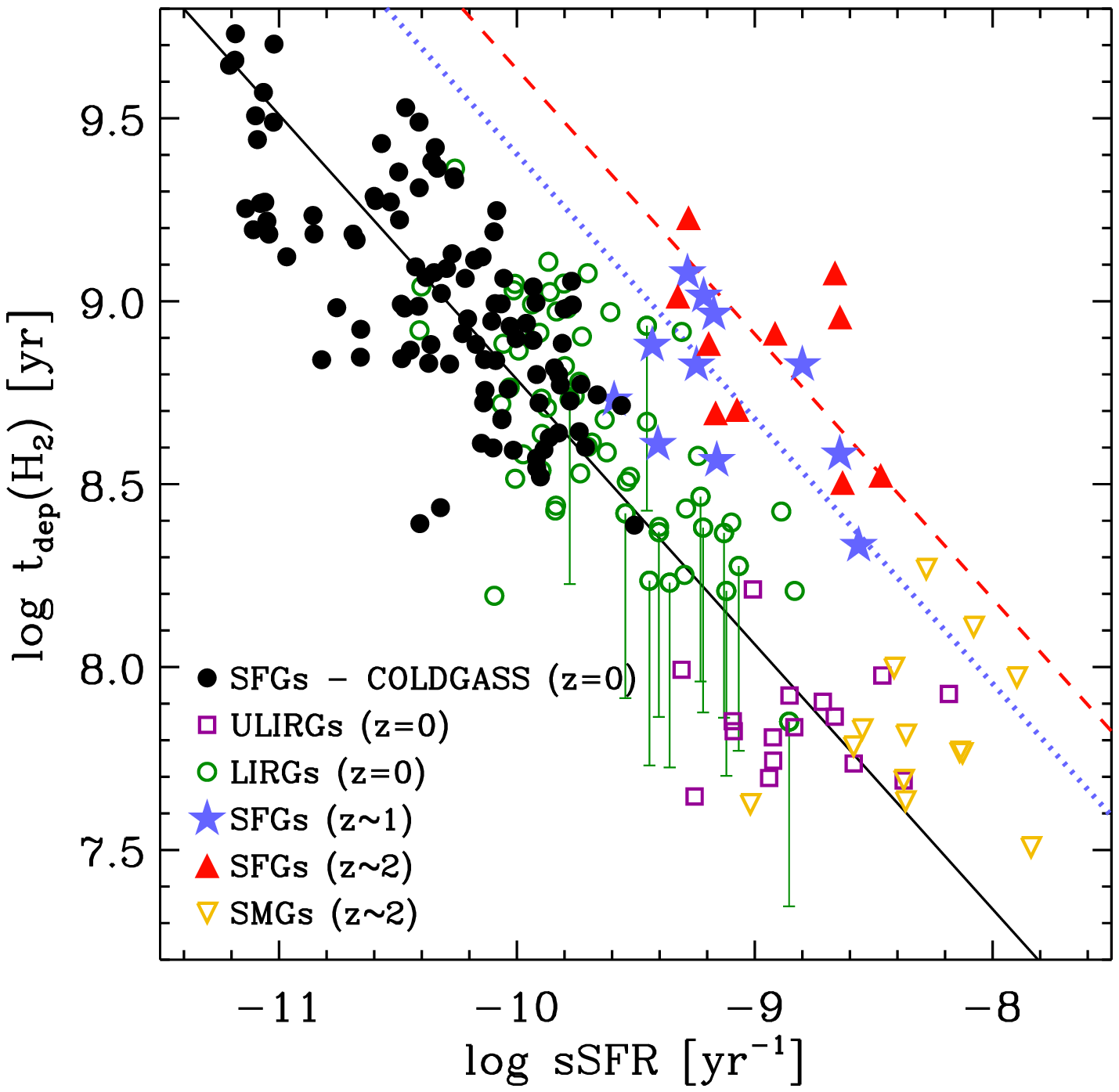}
\caption{Comparison of the depletion time-specific star formation rate relation, 
at $z=$0, 1 and 2, for normal star forming galaxies and extreme systems (local LIRGs/ULIRGs and high-$z$ SMGs).  The best-fit bisector linear relation to the COLD GASS sample, $\log$ \tdep$=(-0.724\pm0.039){\rm sSFR}+(1.54\pm0.41)$, is plotted as a solid black line. This local relation is evolved to $z=$1 and 2 (dotted blue and dashed red lines, respectively) using the observed evolution in sSFR with cosmic time \citep[using the relation of ][]{perezgonzalez08}.   The known offset between the depletion times of normal star-forming galaxies and merging systems 
(local ULIRGs and high $z$ submillimeter galaxies) is seen, but the population 
of local LIRGs is found to bridge this gap.  
A standard conversion factor of $\alpha_{CO}=3.2$\msun\ (K \kms\ pc$^2$)$^{-1}$ 
is adopted for galaxies with $L_{FIR}<10^{12}$\lsun, and a factor 
of $\alpha_{CO}=1.0$\msun\ (K \kms\ pc$^2$)$^{-1}$ is used for the 
ULIRGs and merging systems \citep[these choices are motivated 
in e.g.][]{tacconi10,genzel10}.  The impact of using the smaller 
conversion factor also for the brightest LIRGs ($L_{FIR}>10^{11.75}$\lsun) is shown by a one-sided error bar  
for each individual object. The high redshift data points are 
from \citet{genzel10} and \citet{hainline10}, and the local LIRGs/ULIRGs are 
from \citet{howell10} and \citet{dacunha10}. 
\label{comphighz}}
\end{center}
\end{minipage}
\end{figure*}

While direct \hi\ measurements in individual objects are still only 
possible for nearby galaxies \citep[up to $z\sim0.3$,][]{verheijen07,catinella08}, the detection of CO line emission 
in galaxies at $z>1$ is now achievable and will soon become common 
practice with for example the Atacama Large Millimeter Array (ALMA).   Because of the 
unbiased and homogeneous nature of the sample, COLD GASS will provide 
an ideal local comparison point for these high redshift studies. 
In this section, we take a first look at how the molecular gas
depletion times for the galaxies in our sample compare with
published results at higher redshifts. 

The IRAM Plateau de Bure interferometer is up to now the instrument 
of choice to perform high redshift CO observations. It has been used to 
characterize the molecular gas content of submillimeter galaxies 
\citep[SMGs, e.g.][]{frayer98,downes03,genzel03,neri03,kneib05,tacconi08,bothwell10,engel10}
and normal star forming galaxies at $z>1$ \citep[e.g.][]{tacconi10,genzel10,daddi10}. 
In Figure \ref{comphighz} we compare the position of such systems in the \tdep-sSFR plane.  
To perform this comparison, all data sets need to be put on a common scale. A standard conversion factor of $\alpha_{CO}=3.2$\msun\ (K \kms\ pc$^2$)$^{-1}$ is adopted for galaxies with $L_{FIR}<10^{12}$\lsun, and a factor 
of $\alpha_{CO}=1.0$\msun\ (K \kms\ pc$^2$)$^{-1}$ is used for the ULIRGs and merging systems \citep[these choices are motivated in e.g.][]{tacconi10,genzel10}.  The star formation rates are derived from the infrared luminosity for the LIRGs and ULIRGs, and from the 1.4 GHz flux \citep{chapman05} for the SMGs, using the calibration of the FIR-radio correlation of \citet{magnelli10}. 

For both the high and low redshift populations, merging systems (local ULIRGs
and high-$z$ SMGs), have much shorter depletion timescales
than ``normal'' star-forming galaxies.  These two populations of normal 
and extreme starformers are observed as two distinct branches in 
the Kennicutt-Schmidt relation \citep{genzel10}.  Using the measurements of stellar masses of \citet{howell10} and \citet{dacunha10} for a large sample of local LIRGs, we see that the gap in the relation that extends from the COLD GASS galaxies to the ULIRGs is filled by these galaxies of intermediate infrared luminosities.  

We first focus our attention on the ``normal'' star forming population (filled symbols in Figure \ref{comphighz}). 
The $z=1-2$ galaxies lie at systematically larger values of 
sSFR as compared to our local comparison samples -- this simply reflects the 
well-studied cosmic evolution of the star formation versus
stellar mass relation  \citep[e.g.][]{noeske07,perezgonzalez08,rodighiero10}. 
In Figure \ref{comphighz},  the \tdep-sSFR relation 
derived for the COLD GASS sample is plotted as a solid line. The dotted and dashed
lines show this relation displaced along the x-axis
by factors derived based on \citet{perezgonzalez08} for the evolution of the sSFR 
in blue sequence galaxies out to redshifts of 1 and 2, respectively.
Remarkably, the \tdep-sSFR data points for the $z=1$ and $z=2$ 
galaxies lie almost exactly on these ``shifted'' relations.
The reason why normal high redshift galaxies can have relatively long ($\sim 1$ Gyr) 
molecular gas consumption timescales even though their sSFRs have
increased by nearly an order of magnitude, {\em is because they are
so much more gas-rich than nearby objects} \citep{tacconi10}.
The only galaxies in our local samples that reach sSFR comparable to
the normal high-z star-forming populations are the ULIRGs and some
of the more extreme LIRGs, which are almost all dynamically disturbed
or interacting systems, with much shorter molecular gas depletion times.

Turning now to the full population of galaxies illustrated 
in Figure \ref{comphighz}, we see that when the COLD GASS sample is
combined with samples of low redshift LIRGs and ULIRGs, the \tdep\
versus sSFR relation appears to extend (with much the same slope and even scatter)
over another full 1.5 dex in $\log$ sSFR. Instead of a factor 6 range in
molecular gas depletion time, there is a factor $\sim 50$ decrease from
the quiescent end of the galaxy population where the present-day star formation rate
is only a tenth of its past averaged one, to the most extreme starbursts
where the timescale for making the entire stellar mass of the galaxy
is 1 Gyr or less. There is a hint that the equivalent extreme galaxies at high-$z$, the SMGs, may 
track a different \tdep\ versus sSFR relation -- one that is significantly displaced
with respect to the relation traced by the normal starforming galaxies.  Or alternatively, the slope of the \tdep-sSFR relation may be evolving with redshift.  However, 
testing whether this is the case will require further observations 
of high redshift galaxies, spanning a wider range 
of properties, especially extending to lower (specific) star formation rates.

\section{Summary of observational results}
\label{summary}

A summary of the main empirical results presented in this paper is as follows:

\begin{enumerate}

\item The mean molecular gas depletion timescale for COLD GASS galaxies with
$M_*>10^{10} M_{\odot}$ and detected CO(1-0) line is $1.05\pm0.74$ Gyr.

\item We observe an increase in \tdep\ by a factor of 6 over the stellar mass range
of $10^{10}$ to $10^{11.5} M_{\odot}$, from $\sim 0.5$ to $\sim 3$ Gyr. 

\item Over the same stellar mass range, the atomic gas depletion timescale
remains contant at a value of around 3 Gyr. 
This means that in high mass galaxies, the molecular and atomic gas depletion
timescales are comparable. In low mass galaxies, the molecular gas is being
consumed much more quickly than the atomic gas. 

\item The strongest dependence of \tdep\ is on the specific star formation
rate of the galaxy, estimated using either the UV/optical light (corrected
for extinction) or the 4000 \AA\ break strength. 
This relation can be parametrized as 
$t_{dep}({\rm H_2}) = -0.44(\log {\rm SFR}/M_{\ast} + 10.40)+8.98$, where \tdep\
is in units of $\log$ yr$^{-1}$.   There is also a strong dependence on stellar mass:  \tdep$= (0.36\pm0.07)(\log M_{\ast} - 10.70)+(9.03\pm0.99)$. 

\item We found no significant correlation of the resiuals from these relations
with any of the global galaxy properties we looked at.

\item  The \tdep\ versus sSFR relation extends smoothly with
the same slope from the sequence of ``normal''
star-forming galaxies in our COLD GASS sample through to the population of more extreme
starburst galaxies (LIRGs and ULIRGs), which have \tdep\ $< 10^8$ yr.

\item Normal galaxies at z=1-2 are displaced with respect to the local galaxy
population in the \tdep\ versus sSFR plane, in the sense that they have much
longer molecular gas depletion times at a given value of sSFR. 
This is consistent with the fact that these galaxies are much more gas-rich.
\end{enumerate}

\section{Discussion}
\label{discussion}

In working towards the understanding of the star formation process, 
it probably makes sense to break the problem down into two parts: the conversion
of atomic gas into molecular form, and the subsequent formation of 
stars from this cold dense phase.  One way to  address these issues is 
through resolved studies of nearby galaxy disks, on sub-kpc scales  
\citep[e.g.][]{leroy08} or even down to the scale of giant 
molecular clouds \citep[e.g.][]{gratier10,schinnerer10}.  
Our approach is highly complementary. Only integrated measurements 
are available to us (total \hi, \hmol\ and stellar masses, for example), 
but our COLD GASS sample is large, complete, unbiased, 
and covers a larger region of parameter space than previous studies.

The central result of this paper is that the mean molecular gas depletion timescale (\tdep) increases by a factor of $\sim 6$ across the mass range sampled by COLD GASS galaxies, and that this variation correlates strongly with specific star formation rate, NUV$-r$ colour and stellar mass. Previous surveys were unable to observe this variation or correlation due to the limited dynamic range of their samples.  It should be noted however that the non-universality of \tdep\ is perhaps 
not very surprising.  It is well known that gas  
depletion times are different in extreme starburst galaxies and merging 
systems in the local Universe, as well as submillimeter galaxies 
at high redshifts \citep[e.g.][]{kennicutt98a,riechers07,bouche07,bothwell10}. 
In these galaxies, gas depletion times are significantly shorter, 
with the star formation surface density an order of magnitude larger
 at fixed gas surface density, compared to the normal galaxy 
population \citep{genzel10}.    The novelty of our results lies in showing that \tdep\ varies even within the population of nearby ``normal" star forming galaxies, and in quantifying this variation as a function of a range of fundamental parameters. 

Our work leaves open the intriguing question of what drives the variability in $t_{\rm dep}({\rm H}_2)$ across our sample. In the smaller range of galaxies explored in the THINGS sample, \citet{bigiel08} and \citet{leroy08} find 
a constant molecular gas depletion time.  They do not observe any variation in this quantity with respect to galactocentric radius, Toomre $Q$, or any other galaxy property. This is consistent with the observation by \citet{bolatto08} that molecular cloud properties appear essentially invariant across nearby galaxies with properties similar to those in the THINGS sample.  In starburst galaxies, however, it is clear that $t_{dep}({\rm H_2})$ can be much smaller even than the smallest values found in our sample (e.g. \citealt{gao04}), and in such systems molecular cloud properties are measurably different than in normal galaxies (e.g. in M64, \citealt{rosolowsky05b}). We therefore conjecture that the galaxies in our sample that show significantly reduced \tdep\ are those undergoing minor starbursts, and that the weak correlation between \tdep\ and total stellar mass that we observe arises because smaller galaxies have more bursty star formation histories.  In what follows, we investigate this possible interpretation of the \tdep\ relations, as well as two alternative explanations: quenching in high mass galaxies preventing the molecular gas from efficiently forming stars, and an increasing fraction of molecular gas not traced by the CO(1-0) observations at lower stellar masses.

The first interpretation calls for enhanced star formation efficiency in low mass galaxies due to minor starburst events. This picture is consistent with our result that \tdep\ correlates inversely with specific star formation rate, and that the scatter in the relation is largest at low stellar masses. The starbursts in these galaxies would not be the large ones associated with major mergers, but would instead be weaker ones associated with distant tidal encounters, variations in the intergalactic medium accretion rate, or secular processes within galactic disks. Any process capable of driving gas toward the galactic center and raising the surface density there will suffice, since lower depletion times and altered molecular cloud properties are associated with galactic surface densities above $\sim 100$\msun\ pc$^{-2}$ on both observational and theoretical grounds \citep{krumholz06, krumholz09}.

The second interpretation for the variable \tdep\ invokes the higher mass galaxies, 
and how star formation could be preferentially suppressed in these systems. 
Morphological quenching \citep[e.g.][]{martig09} and
feedback scenarios which prevent the molecular gas from forming stars while
at the same time not destroying the gas, would predict that a consequence of the
quenching would be an increased reservoir of molecular gas.  The specific star formation rate would then inversely correlate with \tdep\ as we observe, and we would expect to see some trend with galaxy morphology (e.g. concentration index, $R_{90}/R_{50}$) which is also seen in Figure \ref{depall}e.  

As a third possible explanation, we ask whether the observed \tdep\ trends may be explained by the fact that the CO(1-0) emission line may not be an accurate tracer of the molecular gas content of the low mass  galaxies in our sample, which may have systematically lower metallicities than the higher mass galaxies \citep{tremonti04}.  \citet{krumholz11} suggest that CO-inferred molecular gas measurements may begin to miss an increasing fraction of the total H$_{2}$ at metallicities as high as half solar \citep[see also][]{bolatto08,leroy11}.  As discussed in Paper I, the COLD GASS sample is limited to galaxies with stellar masses greater than $10^{10} M_{\odot}$, so the {\em global} metallicities of our galaxies are all solar or greater, so this concern is not likely to apply.  One might nevertheless still worry that metallicities in the outer regions of the galaxies are lower, so one might still not account for some fraction of the total molecular content of the galaxy. If metallicity gradients were to correlate with for example stellar mass or stellar mass surface density,  this might create the dependence of $t_{dep}({\rm CO})$ on those quantities.  However, our recent analysis of metallicity gradients using long-slit spectra obtained from MMT (Moran et al., in preparation) indicates that metallicity gradients of the galaxies in our sample are mostly flat and do not correlate strongly with either quantity.

The observed dependence of \tdep\ on several parameters such as stellar mass and specific star formation rate, is then likely caused by a combination of the first two effects, each acting at a different end of the COLD GASS stellar mass range.   In high mass galaxies, star formation would preferentially be quenched by internal processes, while in low mass galaxies, star formation would preferentially be enhanced through mild starburst events. 

The final mystery in this picture is why \tdepHI\ should vary less than \tdep. We note that the invariance of \tdepHI\ must break down in sufficiently strong starbursts -- for example Arp 220 has a star formation rate of $\sim 50$ $M_\odot$ yr$^{-1}$ \citep{downes98}, but an HI mass below $4\times 10^9$ $M_\odot$ \citep{baan87}, giving \tdepHI$ < 100$ Myr. However, galaxies like Arp 220 are rare, and we almost certainly do not have any galaxies quite that extreme in our sample.  For the weaker starbursts we do include, it seems plausible that the majority of the HI resides in a quiescent outer disk that is largely unaffected by whatever process produces the starburst.  
It indeed appears that the conversion of atomic to
molecular gas is, in some situations, a bottleneck in the star formation 
process.  For example, at low stellar mass surface densities 
($\log \mu_{\ast}<8.7$, see Figure \ref{depall}b), all galaxies 
have \tdep\ shorter than the sample mean of 1 Gyr.  These galaxies 
are processing their molecular gas into stars faster 
than they are consuming their atomic gas, due to some internal (e.g. bar instabilities) or external (e.g. tidal interactions) processes triggering minor starburst events, as we speculated above.  
In these galaxies, \hi\ is likely 
to be accumulating in regions where the conversion between atomic and molecular gas 
is less efficient and where this gas cannot participate in the the star formation process, for example in the outer disks of galaxies.  

Why should \hi\ accumulate more effectively in low mass galaxies than in high
mass galaxies? One possibility is that gas from the external environment 
is able to accrete more efficiently in such systems. At a halo mass of greater
than few $\times 10^{11} M_{\odot}$, gas is expected to shock-heat to the virial
temperature of the halo and cooling times become longer \citep[e.g.][]{keres05}.
Once the gas has been accreted, the production of molecular gas may rely
of gas flows that bring gas from the outer disks to the central regions
of the galaxy, where it is able to reach high enough densities to be shielded
from the ambient UV radiation field and form molecules. If such gas flows
are dynamically driven \citep[e.g.][]{chakrabarti09}, one might 
expect them to occur more frequently in high mass galaxies, which are located 
in more massive halos and more crowded environments. In addition, 
feedback processes from AGN may disrupt gas accretion in the most massive dark matter halos.

It is also interesting to speculate on the issue of gas accumulation in high redshift
galaxies. We have seen that the molecular gas depletion times in normal galaxies
at $z=1-2$ are nearly an order of magnitude longer than in local galaxies with
the same specific star formation rates. In  nearby starburst galaxies, the very high
values of sSFR are achieved because dynamical disturbances act to compress the
available interstellar medium  into a very small volume, in which the filling factor of GMCs and
young stars is presumably very high. Although we have not yet been able to demonstrate
this conclusively, our current best guess is that the relation we uncovered between
\tdep\ and sSFR is simply a reflection of increased dynamical stirring of one
sort or another, as one proceeds to more strongly star-forming galaxies.
At the extreme high sSFR end of the sequence, clear evidence for recent merging is always
seen, but at the low sSFR end, the diagnostics of the dynamical effects that
are at work may be
much more subtle. We are currently in the process of analyzing whether
bar-driven inflows, for example, might account for some of the trend internal
to our COLD GASS sample. 

In  high redshift galaxies, however, the high
values of  sSFR are achieved simply because there is  a {\em lot} of molecular gas. 
This leads us to speculate that at high redshifts the star formation bottleneck
is no longer in the conversion of atomic to molecular gas. Indeed, the
flat dependence of $\Omega_{HI}$ with redshift inferred from the abundance of damped 
Lyman alpha systems in high redshift quasar spectra would  appear to rule out
the presence of large HI reservoirs in high redshift galaxies \citep[e.g.][]{hopkins05}. 
Although gas accretion rates are expected to be much larger
at high redshifts, most of the gas must be transformed to molecular
form almost immediately. The star formation bottleneck in high redshift galaxies may
then have more to do with the formation of the proto-stars themselves.

\section*{Acknowledgments}
This work is based on observations carried out with the IRAM 30 m telescope. IRAM is supported by INSU/CNRS (France), MPG (Germany), and IGN (Spain). We sincerely thank the staff of the telescope for their help in conducting the COLD GASS observations. 

RG and MPH are supported by NSF grant AST-0607007 and by a grant from the Brinson Foundation.


\appendix

\section{Star formation rates}
\label{SFRap}

\begin{figure*}
\begin{minipage}{165mm}
\includegraphics[width=165mm]{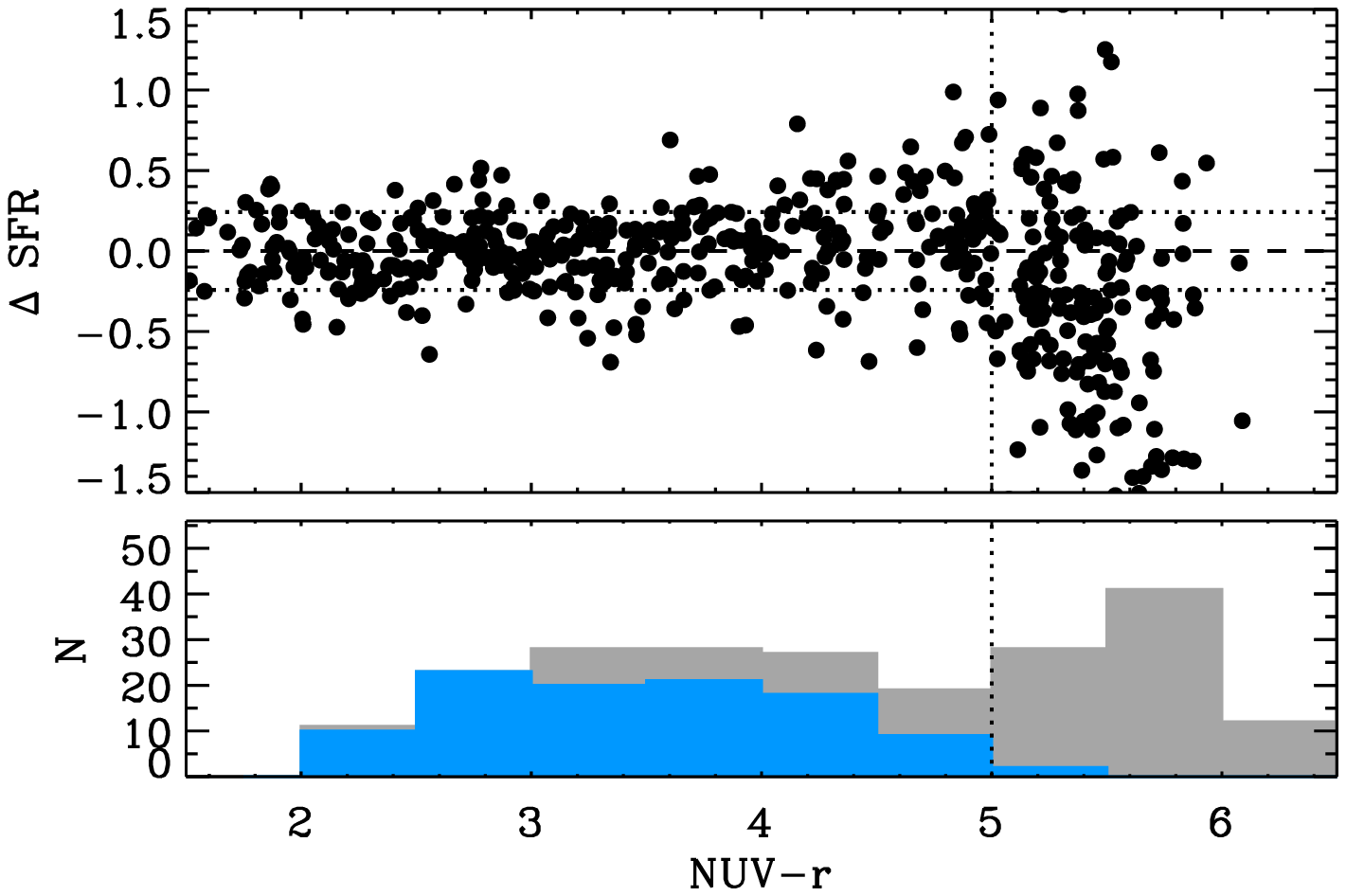}
\caption{Difference in the SFRs measured using the SED-fitting technique and a combination of UV and IR photometry ($\Delta {\rm SFR}=\log ({\rm SFR}_{SED})-\log ({\rm SFR}_{tot})$).  The values of $\Delta {\rm SFR}$ are plotted as a function of \nuvr\ colour (top panel), and the colour distribution of the COLD GASS sample is shown in the bottom panel (in gray the complete sample, in blue the galaxies with CO detection).  Over the colour range where CO detections are possible (\nuvr$<5$), the scatter between the two SFR measurements is only 0.24 dex, with no systematic offset.  \label{compsfr}}
\end{minipage}
\end{figure*}

We use a spectral energy distribution (SED) fitting technique to derive the star formation rate (SFR) for each galaxy in the COLD GASS sample. The \citet{bc03} population synthesis code is used to generate model galaxies with a range of ages, star formation histories, metallicities and dust attenuations and to create a library of model SEDs in 7 bands (FUV, NUV, $u$, $g$, $r$, $i$ and $z$). For each galaxy, we evaluate the goodness-of-fit of each model SED and then estimate its SFR and dust attenuation factor as the $\chi^2$-weighted average of the model parameters.  

This technique is very similar to \citet{salim07} in the way the metallicity, age and star formation history (SFH) priors are set.  There are, however, two main differences compared to their methodology: (1) we only use an exponentialy declining SFH without bursts, because massive galaxies such as those in the COLD GASS sample do not tend to have bursty SFHs, and (2) we use an extinction coefficient ($A_V$) that depends on \nuvr\ colour, after calibration using a reference sample with direct measurement of the SFR through UV and infrared measurements.  The technique and the calibration of $A_V$ are explained in more detail below.  

As pointed out by \citet{salim07} and \citet{wang10}, the derived dust  
attenuation (and thus SFR) is sensitive to the assumed prior  
distribution of $A_V$. A direct estimate of $A_V$ in a galaxy is possible only
if we know its UV through far-IR SED. This is not the case for our sample, 
so we have assumed a different prior $A_V$ distribution for each individual galaxy, constrained by an  
indirect estimate of $A_V$ in the galaxy. Two different methods
of estimating the $A_V$ prior are considered:   
the measured  Balmer decrement from the SDSS spectrum, and a    
combination of D4000 and \nuvr\ colour \citep{johnson07}. The former  
method measures the extinction experienced by stars located
in HII regions \citep{calzetti94} in the  
central 3\arcsec\ of the galaxy, while the latter is a rough estimate of the  
attenuation of the light from stars with ages of a few hundred million years
\citep[see also][]{GASS2}. We denote the $A_V$ measured from these two methods as $A_V$(B) and $A_V$(N), respectively.

We make use of the GMACS (Galaxy Multiwavelenth Atlas From Combined Surveys) sample \citep{johnson07} to tune our choice of the assumed prior distribution of $A_V$ in the model library, and to test the robustness of the SFR derived from the SED fitting. The 70$\mu$m fluxes are converted into total infrared luminosities ($L_{TIR}$), taking into account a correction factor from the ratio of the observed fluxes at 24$\mu$m and 8$\mu$m \citep{boquien10}. We apply the dust model from \citet{meurer99} and the extinction curve from \citet{calzetti94} and calculate $A_V$ from the ratio of $L_{TIR}$ over the luminosity in the FUV ($A_V$(IRX)). We use the \citet{kennicutt98} relation to convert $L_{TIR}$ to SFR (SFR$_{TIR}$) and divide it by a factor of 1.7 to account for the difference between Salpeter IMF and Chabrier IMF. We calculate SFR from obscured FUV luminosity (SFR$_{FUV}$) using the formula from \citet{salim07}.  The total star formation rate for the GMACS galaxies is therefore obtained as SFR$_{tot}$=SFR$_{TIR}$+SFR$_{UV}$, with the latter not corrected for dust extinction.

Different prescriptions for $A_V$ based on the SDSS data are tested against the directly measured values of $A_V$(IRX).  We find that different techniques to measure for $A_V$ minimize best the scatter with $A_V$(IRX) in different \nuvr\ colour intervals, and therefore adopt the following prescription:
\begin{equation}
A _{V}({\rm best}) = \Bigg \{ \begin{array}{ll}
 A_V (B)-0.052({\rm NUV}-r)  & \mbox{if \nuvr $<2.5$}\\
0.5 (A_V(B)+A_V(N)) & \mbox{if $2.5<$\nuvr$<3.5$}\\
A_V(N) & \mbox{if $3.5<$\nuvr$<5.0$}\\
1.5 & \mbox{if \nuvr$>5$}.
\end{array}  \label{AVeq}
\end{equation}
Therefore, when we estimate the SFR for an individual galaxy, we adopt a Gaussian prior distrubtion of $A_V$ based on its observed colour that peaks at $A_V$(best) as given in Equation \ref{AVeq} and has a width of $\sigma=0.15$. We compare SFR from the SED fitting method with the ``true" SFR$_{tot}$ measured from the UV and IR photometry in Figure \ref{compsfr}. The two sets of SFRs are very consistent when \nuvr$<5$, with a scatter of 0.24 dex in their log difference.  The scatter over the full colour distribution is 0.37 dex, because of the low $S/N$ in the GALEX images for the redder galaxies.  This, however, has no influence on the results of this paper, since all COLD GASS galaxies detected in CO have \nuvr$<5$ where the SED-fitting technique produces reliable SFR measurements.  

Another approach to measure SFRs for the COLD GASS galaxies would have been to correct the GALEX FUX flux using $A_V$(best) to infer the total SFR (SFR$_{FUV,cor}$), instead of relying on SED fitting using a model library.  This was the approach taken in \citet{johnson07}, and we test it on the GMACS sample.  The scatter in the log difference of SFR$_{tot}$ and SFR$_{FUV,cor}$ is 0.29 dex for galaxies with \nuvr$<5$.  Over the same colour range, the scatter between SFR$_{tot}$ and SFR$_{SED}$ is 0.24, indicating that information is gained by using the full 7 optical and UV bands from SDSS and GALEX in the determination of the SFRs, if the dust attenuation prior distribution can be well constrained.

\section{Residuals of the molecular gas depletion time relations}
\label{otherparams}

\begin{figure*}
\begin{minipage}{165mm}
\includegraphics[width=165mm]{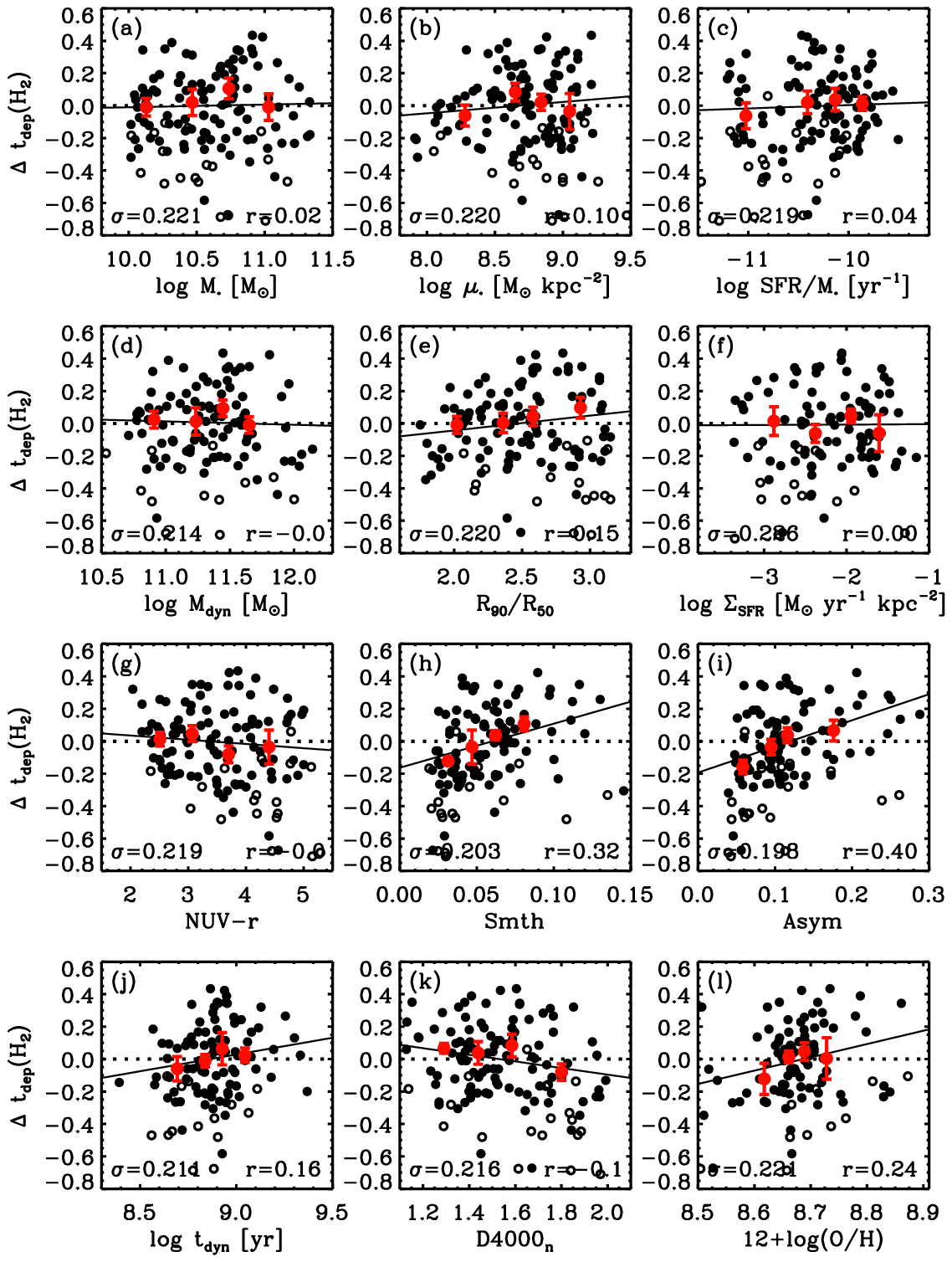}
\caption{Residuals of the sSFR-\tdep\ relation of Figure \ref{depall}c, as a function of 12 different global measurements for the COLD GASS galaxies with secure (filled circles) and tentative (open circles) CO detections.  The red symbols show the mean value of the residuals in equally-populated bins of the $x-$axis parameter.  The correlation coefficient ($r$) of the data in each plane as well as the scatter around the best fit relation ($\sigma$) are given in each case. 
\label{residuals}}
\end{minipage}
\end{figure*}

As explained in Section \ref{tdeprelations} and shown in Figure \ref{depall}, the molecular gas depletion time (\tdep) correlates best with specific star formation rate (sSFR).  To investigate what other physical parameters may be at play in setting this relation, we examine here the residuals of the sSFR-\tdep\ relation as a function of the other global parameters considered in this study.  The (data - fit prediction) residuals are shown in Figure \ref{residuals}, where a positive residual therefore indicates a longer depletion time than average at fixed sSFR.  

The only two parameters upon which the residuals depend are the measures of smoothness and asymmetry (Figure \ref{residuals} h and i).  However, the trends are driven mostly by a small number of points at both low and high values of Smth and Asym, and the amplitude of the trend seen in the binned values is smaller than the overall calibration accuracy of the star formation rates (Appendix \ref{SFRap}) and the uncertainty on the molecular gas masses.  Furthermore, if we were instead to show the residuals of the stellar mass-\tdep\ relation, the trend seen with Smth and Asym is even weaker.  We therefore conclude that there is no significant third parameter dependence among the quantities we have examined.  We will exmine this issue in more detail in a later paper.

\end{document}